\documentclass[pdflatex,sn-mathphys-num]{sn-jnl}

\usepackage{graphicx}%
\usepackage{multirow}%
\usepackage{amsmath,amssymb,amsfonts}%
\usepackage{amsthm}%
\usepackage{mathrsfs}%
\usepackage[title]{appendix}%
\usepackage{xcolor}%
\usepackage{textcomp}%
\usepackage{manyfoot}
\usepackage{booktabs}%
\usepackage{natbib}%
\usepackage{algorithm}%
\usepackage{algorithmicx}%
\usepackage{algpseudocode}%
\usepackage{listings}%
\usepackage{ulem}%
\usepackage{tabularx}%
\usepackage{array}%

\theoremstyle{thmstyleone}%
\theoremstyle{thmstyletwo}%

\theoremstyle{thmstylethree}%

\begin{document}

\title{CONFIDE: Hallucination Assessment for Reliable Biomolecular Structure Prediction and Design}



\author[1]{\fnm{Zijun} \sur{Gao}}
\equalcont{These authors contributed equally to this work.}

\author[3]{\fnm{Mutian} \sur{He}}
\equalcont{These authors contributed equally to this work.}

\author[3]{\fnm{Shijia} \sur{Sun}}

\author[1]{\fnm{Hanqun} \sur{Cao}}

\author[1]{\fnm{Jingjie} \sur{Zhang}}

\author[4]{\fnm{Zihao} \sur{Luo}}

\author[2]{\fnm{Xiaorui} \sur{Wang}}

\author[3]{\fnm{Xiaojun} \sur{Yao}}
\author*[2]{\fnm{Chang-Yu} \sur{Hsieh}}\email{kimhsieh@zju.edu.cn}

\author*[1]{\fnm{Chunbin} \sur{Gu}}\email{guchunbin200888@gmail.com }

\author[1]{\fnm{Pheng Ann} \sur{Heng}}

\affil[1]{\orgdiv{Department of Computer Science and Engineering}, \orgname{The Chinese University of Hong Kong}, \orgaddress{\state{Hong Kong}}}

\affil[2]{\orgdiv{College of Pharmaceutical Sciences}, \orgname{Zhejiang University}, \orgaddress{\city{Hangzhou}, \postcode{310027}, \state{Zhejiang Province}, \country{China}}}

\affil[3]{\orgdiv{Faculty of Applied Sciences}, \orgname{Macao Polytechnic University}, \orgaddress{\state{Macao}}}

\affil[4]{\orgdiv{School of Mechanical and Electrical Engineering}, \orgname{University of Electronic Science and Technology of China}, \orgaddress{\city{Chengdu}, \postcode{611731}, \state{Sichuan Province}, \country{China}}}


\abstract{Reliable evaluation of protein structure predictions remains challenging, as metrics like pLDDT capture energetic stability but often miss subtle errors such as atomic clashes or conformational traps reflecting topological frustration within the protein-folding energy landscape. We present CODE (Chain of Diffusion Embeddings), a self-evaluating metric empirically found to quantify topological frustration directly from the latent diffusion embeddings of the AlphaFold3 series of structure predictors in a fully unsupervised manner. Integrating this with pLDDT, we propose CONFIDE, a unified evaluation framework that combines energetic and topological perspectives to improve the reliability of AlphaFold3 and related models. CODE strongly correlates with protein folding rates driven by topological frustration, achieving a correlation of 0.82 compared to pLDDT’s 0.33 (a relative improvement of 148\%). CONFIDE significantly enhances the reliability of quality evaluation in molecular glue structure prediction benchmarks, achieving a Spearman correlation of 0.73 with RMSD, compared to pLDDT’s correlation of 0.42, a relative improvement of 73.8\%. Beyond quality assessment, our approach applies to diverse drug-design tasks, including all-atom binder design, enzymatic active-site mapping, mutation-induced binding-affinity prediction, nucleic acid aptamer screening, and flexible protein modeling. By combining data-driven embeddings with theoretical insight, CODE and CONFIDE outperform existing metrics across a wide range of biomolecular systems, offering robust and versatile tools to refine structure predictions, advance structural biology, and accelerate drug discovery.}

\keywords{Protein Folding, Topological Frustration, Energy Landscape, AlphaFold3}

\maketitle 

\section{Introduction}\label{sec1}

The advent of deep learning in structural biology, epitomized by AlphaFold3 (AF3), has transformed our ability to predict the three-dimensional structures of biomolecules including proteins, nucleic acids, and their complexes \cite{abramson2024accurate}. AF3 often achieves near-experimental accuracy across a broad range of targets, accelerating progress in structural biology and drug development. Yet, despite these advances, the internal mechanisms by which AF3 generates its predictions, and the reliability of its self-assessed confidence, remain incompletely understood.

AF3 reports a predicted Local Distance Difference Test (pLDDT) score to quantify structural confidence. While pLDDT generally correlates with model accuracy, it fails in critical cases involving complex molecular assemblies or atypical conformations. Similar to other large-scale generative models, AF3 can produce hallucination, i.e. structures with deceptively high confidence but substantial deviations from experimentally determined conformations. This high-confidence/low-accuracy paradox undermines downstream applications, particularly in drug discovery, where misleading structural models may misdirect experimental resources. A deeper understanding of AF3’s generative process is therefore crucial for mitigating hallucinations and improving reliability, especially in data-scarce scenarios.

Classical interpretations of AlphaFold draw on protein-folding energy landscape theory. AlphaFold2 (AF2) was hypothesized to learn an energy function, using coevolutionary information from multiple sequence alignments (MSAs) to identify approximate global minima, followed by refinement through its structure module \cite{roney2022state}. While this framework helps explain how AF2 resolves energetic frustration, it does not account for the additional constraints imposed by protein topology. Protein folding also reflects topological frustration, which arises from chain connectivity, conformational entropy losses, and the need to traverse rugged folding funnels without becoming trapped in local minima \cite{brockwell2000protein,grantcharova2001mechanisms,gunasekaran2001keeping}. These topological constraints profoundly shape folding pathways and increase structural complexity, yet these intricate details of folding process are not well captured by existing data-driven methods.

Building upon these insights, we investigate whether the generative process of the AlphaFold3 series of structure predictors can be understood more comprehensively through the lens of energy landscape theory. Here, we introduce a theoretical framework for interrogating the predictions of the AlphaFold3 series of structure predictors by quantifying topological frustration. Inspired by chain-of-thought reasoning in large language models (LLMs), we formalize the progressive latent representations within the diffusion-based structure module of the AlphaFold3 series of structure predictors as the Chain of Diffusion Embeddings (CODE) \cite{wang2024latent,wei2022chain}. We hypothesize that these progressively refined embeddings capture long-range inter-residue relationships reflective of backbone constraints and topological complexity \cite{peters2018dissecting}. As mentioned above, these subtle structural constraints cannot be easily recognized by more localized metrics such as pLDDT. To validate this, we conducted proof-of-concept experiments demonstrating that CODE strongly correlates with protein folding rates driven by topological frustration, achieving a correlation of 0.82 compared to pLDDT’s 0.33 (a relative improvement of 148\%).

Recognizing the complementarity  between CODE and pLDDT, we develop CONFIDE (CONFIdence-integrated Diffusion Embedding), a self-evaluating metric that unifies energetic (pLDDT) and topological (CODE) perspectives to provide a more robust assessment of structural predictions. Across diverse benchmarks—including ternary complexes, flexible and rigid proteins, and other challenging datasets—CONFIDE achieves superior performance compared to pLDDT, with particularly notable improvements in molecular-glue complexes (31\% Spearman correlation enhancement). Furthermore, in downstream applications, CONFIDE operates in a fully unsupervised manner to guide binder design, enzyme catalytic site identification, mutation-induced affinity prediction, and inhibitor/nucleic acid aptamer screening, achieving excellent results across all tasks. Notably, CONFIDE improves the success rate of binder design for IAI by 13\%. Additionally, CONFIDE accurately detects affinity changes from resistance mutations in the BTK protein against Fenebrutinib, with a Spearman correlation of 0.83, far exceeding pLDDT’s negligible 0.03.

By integrating embedding-based reasoning with theoretical insights into protein folding, CODE and CONFIDE provide a unified, interpretable framework for detecting structural hallucinations and improving model reliability. 
Addressing a key limitation in diffusion-based 3D generative modeling—widely used in structural biology, materials science, and engineering \cite{zeni2023mattergen}—this framework universally applies to computational models with progressive structural encoding, ensuring robust evaluation of topological rationality across diverse domains.
\section{Results}\label{sec2}

\subsection{Overview}
\begin{figure}[htbp]
    \centering
    \includegraphics[width=1\textwidth,height=0.86\textheight]{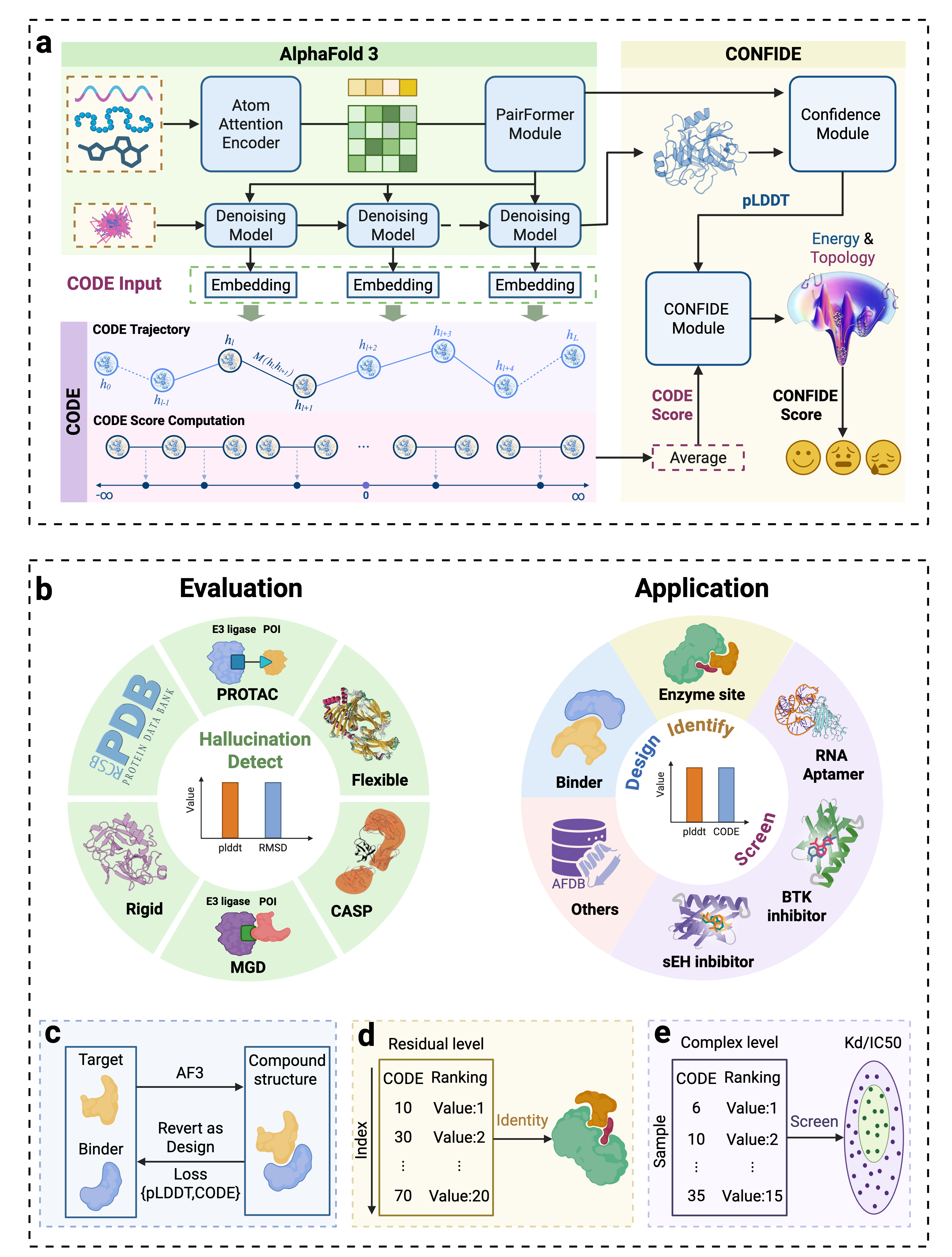}
    \caption{\textbf{Overview of CODE and CONFIDE.} (a) The computational process of the CODE module and the CONFIDE module. First, the sequence of any biological molecule is input to the Alphafold3 series model to obtain the PairFormer embedding. Then, the embedding guides the noise to generate the corresponding structure. We construct the embedding change trajectory in the Denoising Module and calculate the CODE score to characterize topological frustration. The generated structure is used as the input of the Confidence Module to obtain the confidence score to represent energy frustration. Finally, the CONFIDE Module comprehensively considers the two frustrations and analyzes to obtain a more comprehensive CONFIDE score. (b) Shows the application scenarios of CODE and CONFIDE. In the hallucination detection of structure prediction, we verified 6 challenging datasets covering almost all biological molecules. We also expanded the evaluation to five application scenarios covering three types of tasks: binder design, enzyme site detection, and complex screening. (c) Shows the workflow of CODE application to hallucination-based binder design. (d) Shows the schematic diagram of CODE for residue level site detection. (e) Shows the schematic diagram of CODE/CONFIDE for virtual screening}
    \label{framework}
\end{figure}

Inspired by the topological frustration theory in protein folding, we reformulated the diffusion embedding trajectories of the AF3 series structure predictors into CODE, a metric that captures topological frustration overlooked by the conventional confidence metric pLDDT. We propose CONFIDE, a unified framework integrating topological frustration and pLDDT-represented energetic frustration to comprehensively characterize the protein folding energy landscape (see Figure \ref{framework} for a schematic of the framework). To establish CODE’s significant correlation with topological frustration, we conducted proof-of-concept experiments on single-domain proteins with minimal energetic frustration. These experiments demonstrated a strong correlation between CODE and topological frustration, a weak correlation between pLDDT and topological frustration, and weak coupling between CODE and pLDDT, providing information gain-based interpretability for CONFIDE’s robustness over individual metrics.

We categorized the applications of CODE and CONFIDE into two classes: hallucination detection and downstream applications. For hallucination detection, we evaluated ternary complexes, flexible and rigid proteins, and benchmark datasets (PDB test and CASP15). Comprehensive assessments of classification performance and correlation underscored CONFIDE’s superior performance, with improvements explained from a topological frustration perspective based on atomic clash rates. In downstream applications, CODE and CONFIDE were applied to de novo binder design, enzyme active site identification, drug resistance mutation prediction, and virtual screening. These applications highlight the broad potential of CODE and CONFIDE as unsupervised self-evaluation tools, offering biologically reasonable interpretations (e.g., steric hindrance explaining drug resistance mutations) and enhanced design performance (e.g., generating more stable binder structures).

\begin{figure}[htbp]
    \centering
    \includegraphics[width=1\textwidth]{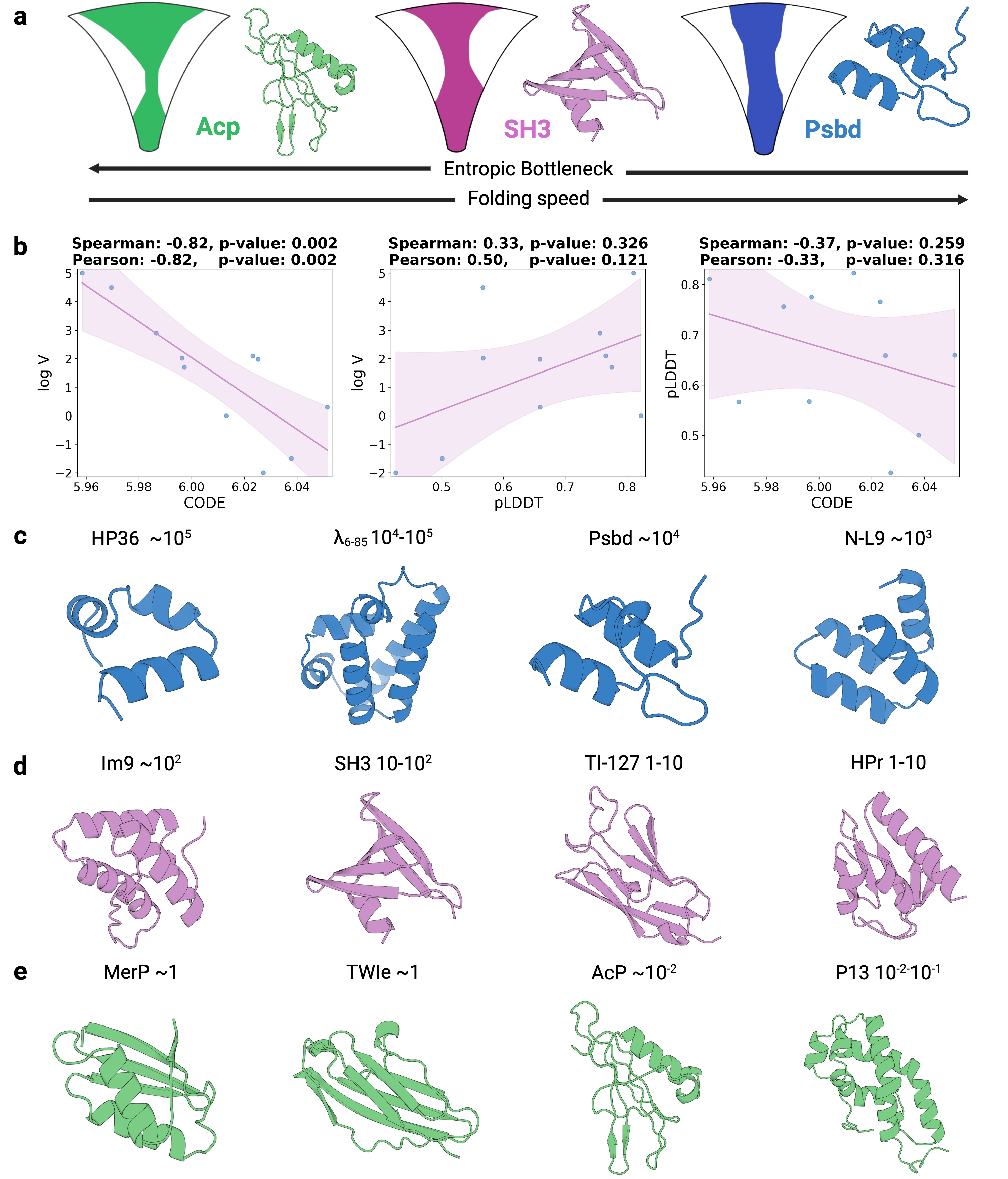}
    \caption{\textbf{CODE implicitly encodes the protein folding dynamics mediated by topological frustration.} (a) Three proteins with different orders of magnitude of folding rates are shown. AcP has mild mixing entropy early in the folding process. A large bottleneck region slows the folding of this protein around the transition-state barrier. SH3 is a slower folding protein and shows little mild mixing entropy initially, and then a small bump appears in the route measure curve. Psbd is a very fast folding protein and shows almost constant mixing entropy measure throughout folding. (b) Correlation analysis among CODE, pLDDT and log V with linear regression fits and 95\% confidence intervals. The first figure shows a strong Spearman correlation between CODE and log V. The second one shows a weak Spearman correlation between pLDDT and log V. The third one shows that there is only a weak correlation of -0.37 between CODE and pLDDT, indicating that they depict the energy landscape of protein folding from different perspective. (c) Protein structures with fast folding rates, showing their names and specific folding rates. (d) Protein structures with intermediate folding rates, showing their names and specific folding rates. (e) Protein structures with slower folding rates, showing their names and specific folding rates.  }
    \label{topo}
\end{figure}


\subsection{Topological Frustration-Mediated Interplay between CODE and Protein Folding Kinetics}

Experimental evidence, together with comparisons to theoretical models, shows that proteins are robust folding entities: they can reach their native conformation under a wide range of conditions and preserve that fold even after multiple mutations in their amino acid sequences \cite{itzhaki1995structure,bryngelson1995funnels,kuhlman2000native,bryngelson1987spin}. For small (single-domain) proteins, experiments further reveal that evolution has selected sequences whose energetic frustration on the free-energy landscape is so low that sensitivity to any particular mutation is the exception rather than the rule. Two observable consequences follow: (i) these proteins fold very rapidly on the microsecond-to-second time scale under typical laboratory conditions and (ii) the structural details of their folding mechanisms are governed mainly by what is commonly called topological frustration \cite{brockwell2000protein,grantcharova2001mechanisms,gunasekaran2001keeping}.

Topological frustration effectively describes the roughness of the folding landscape that arises from the interplay between chain connectivity and energetic bias toward the native state \cite{chavez2004quantifying}. More precisely, this ruggedness results from a heterogeneous loss of conformational entropy associated with the formation of partially folded structures on the free-energy landscape. For proteins in which the heterogeneity of conformational entropy far exceeds that of energy, the principal folding pathways on the free-energy surface are tightly constrained and shaped by the protein’s topology \cite{plotkin2002structural,plotkin2000investigation}.

To more intuitively explain how topological frustration affects the protein folding process and ultimately reflects on the folding rate, we provide three specific protein cases spanning different orders of magnitude of folding rates which is shown in Figure \ref{topo}a. AcP has mild mixing entropy early in the folding process so the protein may sample a large volume of configuration space. A large bottleneck region slows the folding of this protein around the transition-state barrier. Diffuse structure flickers between conformations before the transition-state region, then the formation must become more ordered.  Unfolded AcP may sample a large region of the configuration space available to it before it manages to pass through the narrow bottleneck that leads to the native state. These effects cause AcP to be one of the slowest two-state proteins. SH3 is a slower folding protein and shows little mixing entropy initially (similar to AcP), and then a small bump appears in the entropy measure curve. This indicates that at that time some portion of the protein is more likely to be structured relative to the rest of the protein. These contacts may need to be in place for SH3 to continue folding. After this short region is crossed, the rest of the curve is quite uniform and low. This protein never has its configuration space strongly reduced as it searches for its native state. The overall shape of the curve is similar to AcP, although the entropic bottleneck is smaller for SH3. Psbd is a very fast folding protein and shows almost constant mixing entropy measure throughout folding. The folding process from the unfolded state to the folded state is accompanied by a very slow decrease in the width of the sampled configuration space.

We conducted proof of concept tests of the link between CODE and folding-kinetics framework on a database of protein models that are completely free of energetic frustration \cite{plaxco2000topology,chavez2004quantifying}. Proteins were selected whose experimentally measured folding rates span more than six orders of magnitude and whose overall fold topologies and secondary-structure compositions are diverse. The final data set consists of 16 two-state, single-domain proteins ranging from 36 to 115 residues in length. Structural information and detailed information on folding rates of selected proteins can be found in Supplementary Information \ref{topo_dataset}. For detailed calculation methods of pLDDT and CODE, please refer to Section Methods \ref{method} and Supplementary Information \ref{compute}. Finally, we calculate the correlation coefficients of pLDDT and CODE with respect to the natural logarithm of the folding rate.


As shown in Figure \ref{topo}b, CODE and the protein folding rate log V have a strong spearman correlation of -0.818, with a p-value of 0.002, while pLDDT has only a weak spearman correlation of 0.327, with a p-value of 0.326. This result suggests that CODE likely reflects topological frustration during protein folding, supported by a strong Spearman correlation with a statistically significant p-value. Protein folding is a process from a high-entropy unfolded state to a low-entropy native state, involving multiple intermediate states. Topological frustration may cause proteins to fall into local energy minima in the folding funnel, prolonging the folding time. High values of CODE may correspond to complex folding funnels containing more or deeper local traps. pLDDT, on the other hand, reflects more the energy stability of static structures rather than the dynamic process of folding dynamics. The folding rate depends not only on the stability of the final structure, but also on the folding path, transition states, and topological frustration. The weak correlation of pLDDT suggests that it may not be able to effectively capture these dynamic factors. Furthermore, as shown in the right side of Figure \ref{topo}b, the low correlation of -0.372 between CODE and pLDDT with a p-value of 0.259 suggests that they are a set of features that tend to be weakly orthogonal and can be used together to provide a more comprehensive analysis of the folding process, where CODE captures topological frustration and pLDDT provides energy stability information.

\subsection{Mitigating Hallucination Across Diverse Systems}
Modern structure predictors have essentially reached experimental resolution for single‐chain proteins, yet they still display a striking limitation in biologically realistic settings--the phenomenon we term "structural hallucination" \cite{orr2025improved,rathkopf2025hallucination}. We define a hallucinated structure as one for which the model's internal confidence (pLDDT) is high, while the all-atom RMSD to the corresponding experimental structure remains unacceptably large. This high-confidence/low-accuracy mismatch can mis-rank drug-discovery campaigns, diverting resources toward spurious targets or lead compounds \cite{hong2025good,desai2024review}. We introduce CODE and CONFIDE, an evaluator that operates without external labels and is therefore naturally suited to data-poor or entirely novel protein families. To probe its application breadth, we benchmark CODE across three challenging regimes: (1) PROTAC and molecular-glue ternary complexes, which demand simultaneous assessment of the E3 ligase, the protein of interest and the small molecule, and in which current networks struggle with the dynamic nature of these multi-component assemblies, often overestimating small-molecule interfaces due to insufficient training data on such complex systems; (2) flexible proteins with intrinsically disordered regions or multiple conformational states that present challenges as current methods are optimized for single, well-folded structures and cannot adequately capture the ensemble nature of these dynamic systems and (3) a general benchmark comprising PDB Test and CASP15 targets, spanning membrane, nucleic acid binding and large multimeric assemblies in which data bias precipitates hallucination of unseen structures. We also evaluated the performance on rigid proteins. The detailed experimental analysis will be provided in the Supplementary Information \ref{srigid} together with the results of PDB Test and CASP15 in the Supplementary Information \ref{scasp}.

CODE adds an intrinsic quality evaluation layer to any structure predictor, filtering hallucinated structures without additional supervision. Coupled with high-throughput experimental validation in the future, CODE is poised to (i) accelerate PROTAC/molecular-glue optimization by removing erroneous conformers early in the design--make--test loop and enabling construction of large, trustworthy datasets, and (ii) enhance the next generation of folding networks by serving as a self-distillation weight or adapter-gating signal that supplies real-time uncertainty feedback. Together, these advances offer a systematic remedy for structural hallucination and provide a more reliable structural foundation for AI-driven drug discovery.

\subsubsection{Performance in Ternary Complex Structure Prediction}

\begin{figure}[htbp]
    \centering
    \includegraphics[width=1\textwidth]{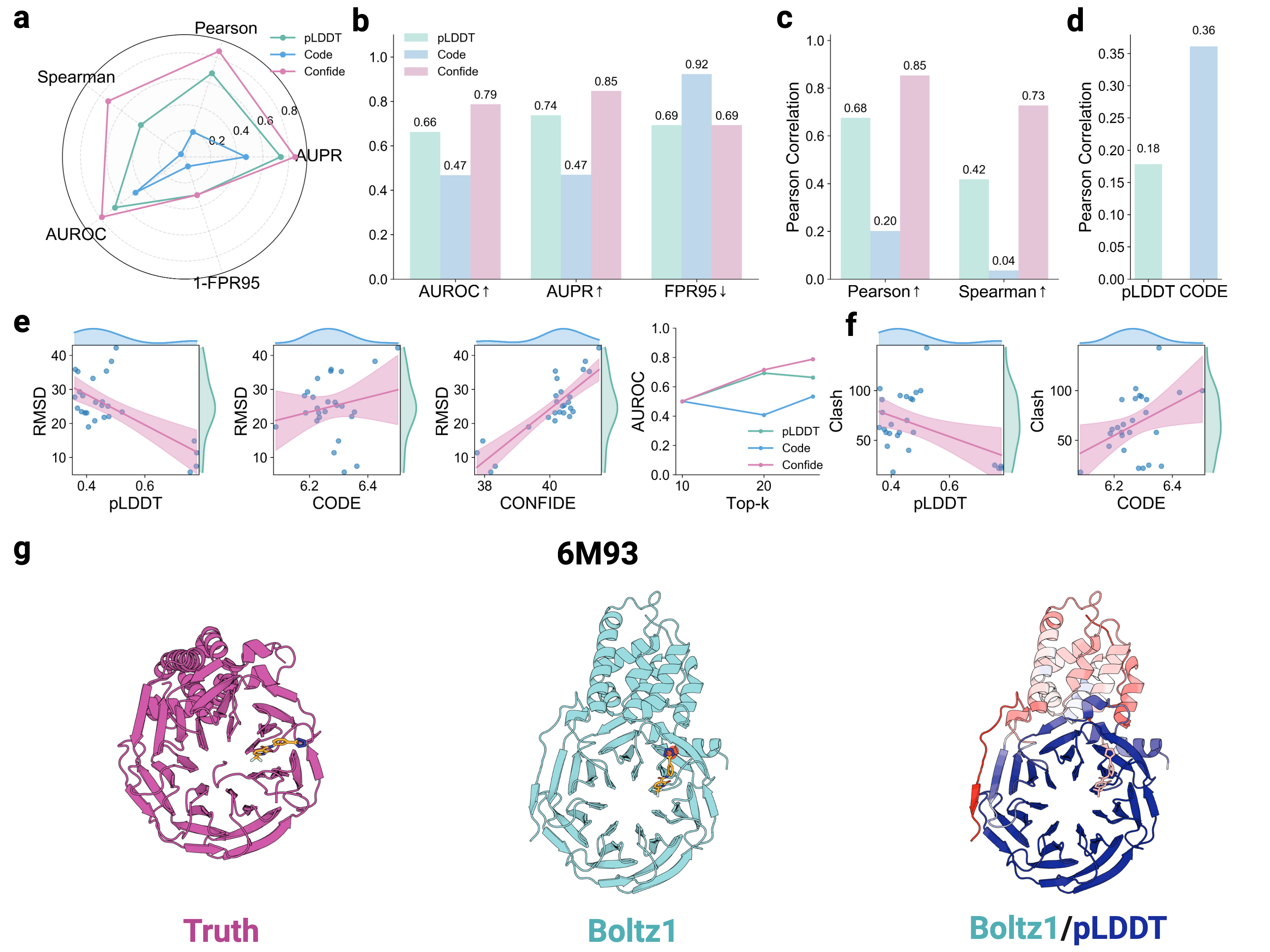}
    \caption{\textbf{Performance analysis on ternary complex structure prediction.} (a) Radar chart of the five evaluation metrics of three scores (pLDDT, CODE and CONFIDE). (b) Histogram of classification evaluation metrics of three scores. (c) Histogram of correlation evaluation metrics of three scores. (d) Spearman correlation between the number of atomic conflicts and pLDDT/CODE. The ternary complex structure predicted by Boltz1 has physically abnormal local conformations, such as steric conflicts caused by atoms being less than 1.5 angstroms apart in space. CODE can capture the high topological frustration caused by these conformations, thus significantly improving the performance of both systems compared to pLDDT.   (e) Scatter plots of the fit and distribution of pLDDT, CODE and CONFIDE with RMSD, including linear regression lines with 95\% confidence intervals, as well as AUROC for different classification thresholds. (f) Scatter plots of the fit and distribution of pLDDT and CODE with the number of atom clashes, including linear regression lines with 95\% confidence intervals. (g) The true structure, predicted structure, and predicted structure colored by pLDDT of 6M93 in MGD are shown from left to right. The red balls in the middle figure indicate atomic conflicts.}
    \label{mgd}
\end{figure}

Targeted protein degradation (TPD) is a revolutionary therapeutic strategy that not only inhibits or activates proteins, but directly eliminates pathogenic proteins \cite{zhao2022targeted,schapira2019targeted}. At present, PROTAC is developing rapidly as one of the main methods of TPD, with more than 20 candidate drugs entering clinical trials and nearly 60 in preclinical development. PROTAC uses the ubiquitin-proteasome system to achieve targeted protein degradation. They promote the formation of a ternary complex between E3 ligase and target protein (POI), resulting in ubiquitination and subsequent degradation of the target protein \cite{bekes2022protac}. Molecular glues are another important class of protein degraders that promote protein degradation by directly binding to and stabilizing the interaction between E3 ligase and substrate protein \cite{toriki2023rational,dewey2023molecular}. Unlike PROTAC, molecular glues are usually small molecule compounds that do not require a linker structure and can induce proteins that do not interact with each other to form stable degradation complexes. Well-known examples include immunomodulatory drugs such as lenalidomide, which can recruit specific substrate proteins to the CRBN E3 ligase complex for degradation \cite{jan2021cancer}.

With the continuous increase in TPD targets and mechanisms, it is essential to develop computational tools to optimize compound design. Current computational methods include molecular dynamics simulation, molecular docking, and machine learning models \cite{xie2023elucidation,zhang2022machine,garcia2022designing,nori2022novo,chen2024interpretable}. Whether it is PROTAC or molecular glue system, the structure prediction of ternary complexes faces great challenges. First, these complexes involve complex interactions between multiple protein components, and their binding modes and conformational changes are difficult to accurately model. Second, the formation of ternary complexes is often a dynamic and transient process, and traditional static structure prediction methods are difficult to capture this time-dependent molecular behavior. In addition, different E3 ligases have different structural characteristics and binding preferences, and the diversity of target proteins further increases the complexity of prediction. For PROTAC, the length, flexibility and chemical properties of the linker significantly affect the stability and geometric configuration of the ternary complex. For molecular glue, it is necessary to accurately predict how small molecules bind to E3 ligases and substrate proteins at the same time and stabilize the interaction interface between them. Most importantly, there is currently a lack of sufficient experimental structural data to train and verify computational models, which makes data-driven structure prediction methods face serious data scarcity problems.

Alphafold3 series structure prediction models have obvious "hallucination" phenomenon in this system. We systematically evaluated the correlation of pLDDT, CODE and CONFIDE with RMSD on PROTACFOLD \cite{erazo2025enhancing}. The PROTACFOLD dataset is derived from 48 known PROTAC-mediated complexes and 33 molecular glue-mediated complexes in the PDB. Detailed dataset information can be found in the Supplementary Information \ref{ternary_dataset}. For detailed calculation methods of pLDDT and CODE, please refer to Section Methods \ref{method} and Supplementary Information \ref{compute}. Finally, we screened all combinations of integer coefficients from -10 to 10 to weight CODE and pLDDT to obtain CONFIDE, and calculated the correlation coefficient under the optimal combination. 

In Figure \ref{mgd}a, we first presented an overall comparison of the performance of the three scores in classification and correlation tasks, observing that CONFIDE demonstrates a significant advantage over pLDDT or CODE alone, achieving comprehensive superiority. In the classification task, CONFIDE improved the AUROC by 0.13 and the AUPR by 0.11 compared to pLDDT which is shown in Figure \ref{mgd}b. In the correlation analysis, CONFIDE achieved a Pearson correlation of 0.85, a 25\% improvement over pLDDT, and a Spearman correlation of 0.73, a 74\% improvement compared to pLDDT's 0.42 which is shown in Figure \ref{mgd}c. In Figure \ref{mgd}e, we displayed the distribution of the three scores across all samples along with their fitted curves, as well as the trend of AUROC changes with different top K selections.

Furthermore, we deeply elucidate the underlying mechanisms for CONFIDE's significant advantages from the theoretical perspective of topological frustration. We observed that in the predicted structures of Boltz1, there were numerous physical anomalies, namely atomic clashes. We define an atomic clash as a spatial distance between atoms of less than 1.5 Å. Such folding leads to a significant increase in topological frustration, which is not reflected in pLDDT scores. To illustrate this phenomenon, we presented a representative molecular glue as a case study (PDB ID: 6M93) in Figure \ref{mgd}g. The leftmost image shows the true crystal structure. The middle image depicts the Boltz1-predicted structure, with atomic clash regions marked by red spheres. The rightmost image colors the predicted structure based on pLDDT scores, with blue indicating high confidence and red indicating low confidence. To some extent, pLDDT can reflect the quality of structural folding, particularly by assigning notably low scores to disordered or loop regions, indicating that pLDDT can capture this type of energy frustration. However, pLDDT is ineffective in detecting atomic clashes. In contrast, atomic clashes significantly increase topological frustration, enabling CODE to capture such structural prediction inaccuracies. We statistically quantified CODE's advantages. Figure \ref{mgd}d shows the Pearson correlation coefficients of pLDDT and CODE with the number of atomic clashes, where CODE achieved a correlation of 0.36, a 100\% improvement over pLDDT’s 0.18. Figure \ref{mgd}f illustrates the specific distribution and fitted curves. These significant improvements demonstrate that CODE captures critical complementary information related to topological frustration, providing substantial benefits for evaluating the structural quality of ternary or even higher-order complexes.

\subsubsection{Assessment of Flexible Proteins}

\begin{figure}[htbp]
    \centering
    \includegraphics[width=1\textwidth]{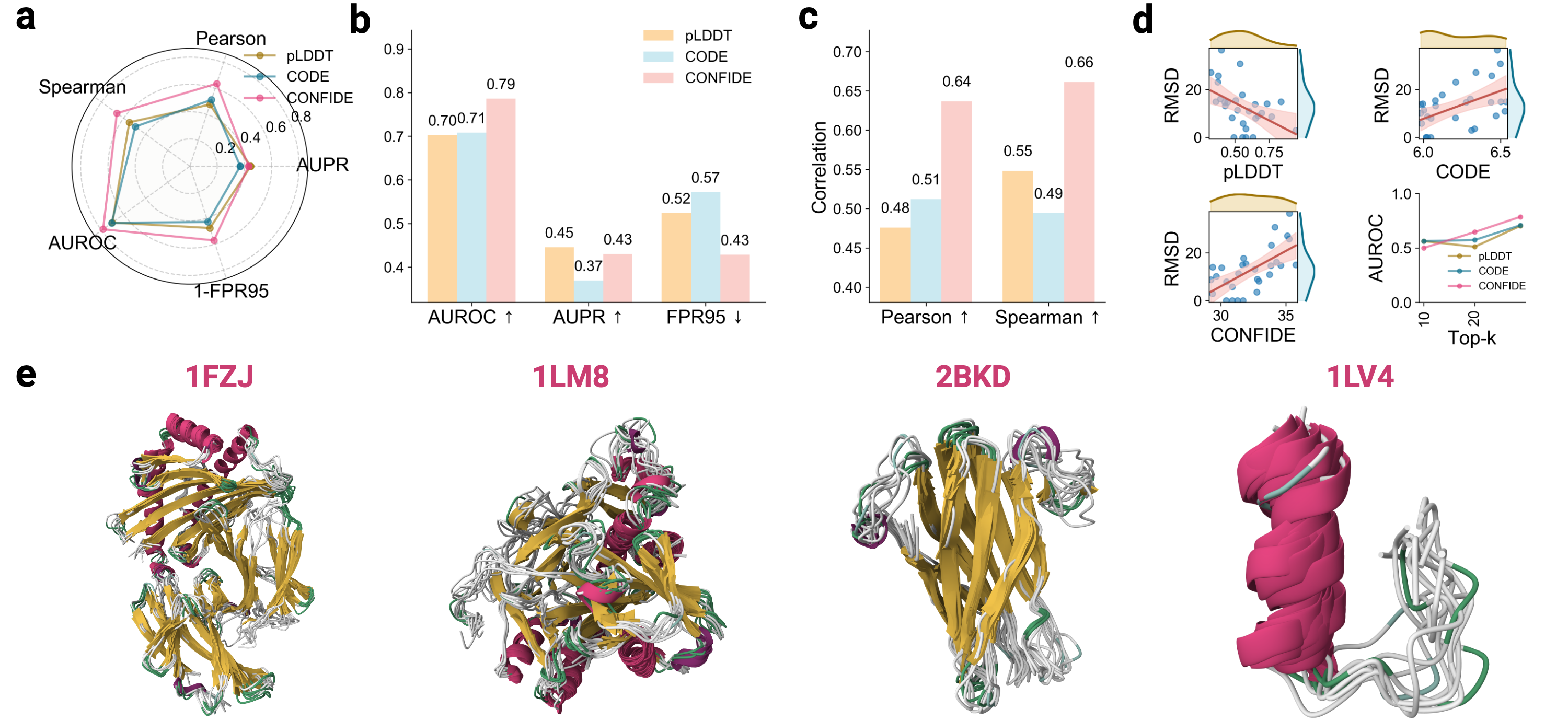}
    \caption{\textbf{Performance analysis on flexible protein structure prediction.} (a) Radar chart of the five evaluation metrics of three scores (pLDDT, CODE and CONFIDE) on flexible protein. (b) Histogram of classification evaluation metrics of three scores on flexible protein. (c) Histogram of correlation evaluation metrics of three scores on flexible protein. (d) Scatter plots of the fit and distribution of three scores on flexible protein with RMSD, featuring linear regression fits with 95\% confidence intervals, as well as AUROC for different classification thresholds. (e) The structures of four flexible proteins efficiently simulated using CABS-flex. The structure is presented in the form of an ensemble of all-atom structures, with different colors representing different secondary structures: yellow for sheet, green for turn, red for helix, and gray for coil. }
    \label{felexible}
\end{figure}

The flexible disordered region gives the same protein the ability to respond quickly to the environment, achieve multifunctional regulation, and phase separation \cite{karshikoff2015rigidity,schlessinger2005protein,schlessinger2006profbval}. The difficulty of flexible proteins lies in their essential conformational heterogeneity and dynamic exchange from milliseconds to seconds. Traditional "single structure" output fails, and sampling must span long time scales and explicitly consider conditions such as pH, ions, and chaperone proteins \cite{meller2023predicting}. At the same time, they are limited by the high cost, noise, and lack of unified evaluation indicators of high-throughput data such as NMR, FRET, and SAXS.

To evaluate the ability of CODE and CONFIDE to detect implicit hallucinations in flexible conformation prediction tasks, we screened 30 disordered proteins with liquid-liquid phase separation from PDB as flexible protein datasets. Detailed dataset information can be found in the Supplementary Information \ref{flexible_dataset}. For detailed calculation methods of pLDDT and CODE, please refer to Section Methods \ref{method} and Supplementary Information \ref{compute}. Finally, we compare the performance of confidence score, CODE and CONFIDE in terms of pearson and spearman correlation coefficient. The results are shown in Figure  \ref{felexible}.

In Figure \ref{felexible}a, we first presented an overall comparison of the performance of the three scores in classification and correlation tasks, observing that CONFIDE demonstrates a significant advantage over pLDDT or CODE alone, achieving comprehensive superiority. In the classification task, CONFIDE improved the AUROC by 0.09 and reduced the FPR95 by 0.09 compared to pLDDT which is shown in Figure \ref{felexible}b. In terms of Pearson correlation, CONFIDE achieved 0.64, a 33.3\% improvement over pLDDT’s 0.48. For Spearman correlation, CONFIDE reached 0.66, a 20\% improvement compared to pLDDT’s 0.55 which is shown in Figure \ref{felexible}c. In Figure \ref{felexible}d, we displayed the distribution of the three scores across all samples along with their fitted curves, as well as the trend of AUROC changes with different top K selections. Figure \ref{felexible}e shows four flexible proteins efficiently simulated using CABS-flex \cite{wroblewski2025cabs}. The structure is presented in the form of an ensemble of all-atom structures, with different colors representing different secondary structures: yellow for sheet, green for turn, red for helix, and gray for coil.


\subsection{Diverse Applications in Structural Biology and Drug Discovery}
Our application is organized into three levels, capturing the broad scope of modern structure-guided research. (i) At the design level, we integrate CODE with a hallucination-based sequence design model \cite{pacesa2024bindcraft,cho2025boltzdesign1} to generate small molecule binders, demonstrating that CODE can steer de novo protein design toward topologically coherent and functionally competent folds. (ii) At the site prediction level, CODE operates zero-shot to identify catalytic and active residues in enzymes, enabling rapid functional annotation of structural data \cite{wang2024multi}. (iii) At the complex prediction level, CODE ranks protein-ligand and protein-RNA complexes, identifying drug resistance in different BTK protein mutations \cite{wang2022mechanisms}, quantifying inhibitor affinities in congeneric sEH series \cite{shen2012discovery} and screening GFP-binding RNA aptamers \cite{huang2024protein}. These applications guide hit-to-lead optimization and anticipatory resistance management. Spanning enzymes, kinases, metabolites, drug-like compounds, and structured nucleic acids, CODE’s chemistry-agnostic nature, lack of dependence on retraining backbone models, and mechanistically interpretable scoring unify energetic and topological insights. This breadth—from binder design to functional annotation and resistance prediction—positions CODE as a versatile and high-impact tool for structural biology and drug discovery.

\subsubsection{Improving Binder Design Capabilities}

\begin{figure}[htbp]
    \centering
    \includegraphics[width=1\textwidth]{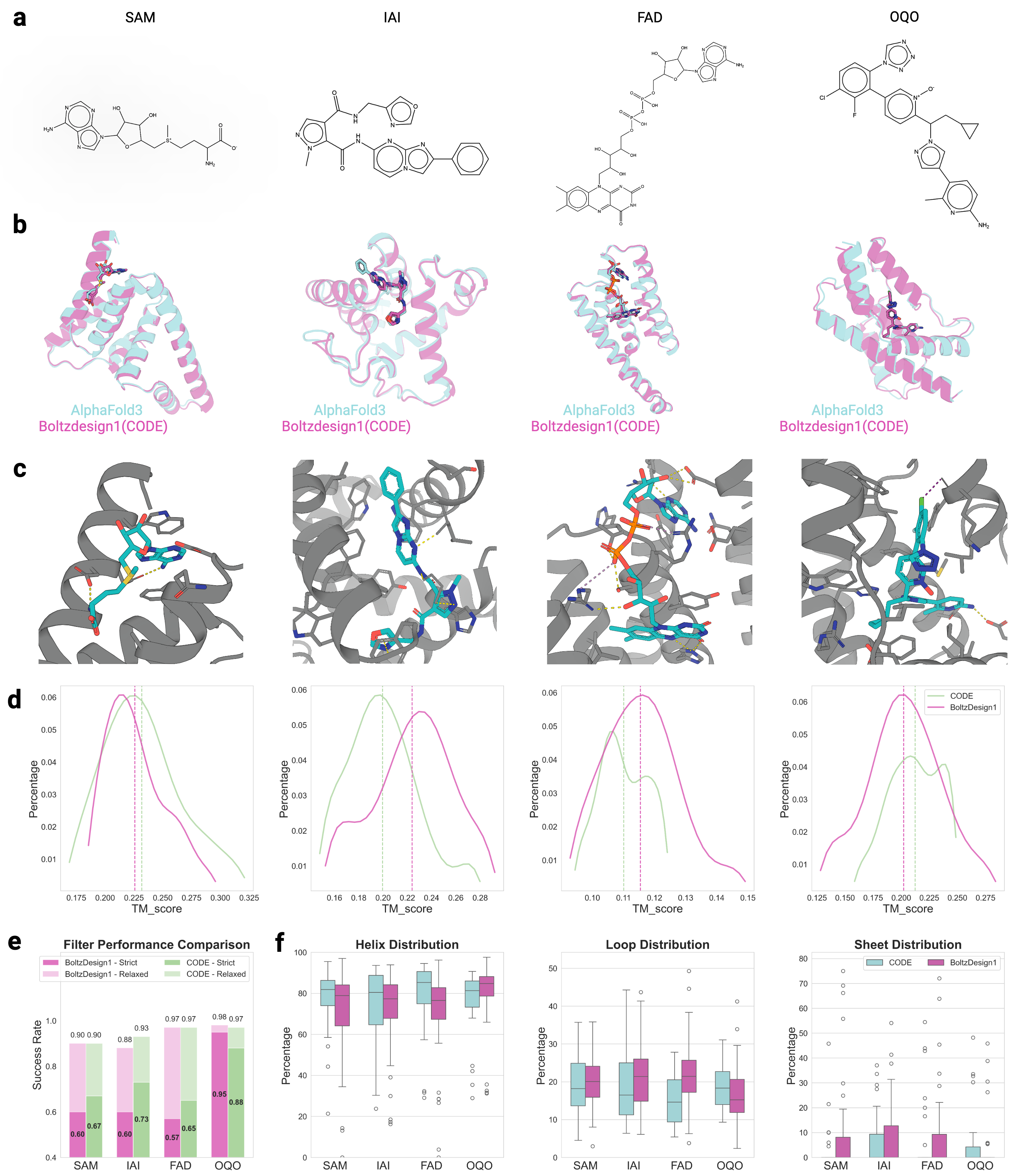}
    \caption{\textbf{CODE guides BoltzDesign1 to generate diverse small molecule binders with high designability and AF3 success rates.}
(a) Chemical structures of four target small molecules used in BoltzDesign1: SAM, IAI, FAD, and OQO. (b) Representative examples of designed binders for these four target molecules, with purple structures showing CODE-optimized designs and green structures showing AlphaFold3 predictions based on the designed sequences. The high structural consistency demonstrates CODE's ability to successfully design binders for small molecule ligands. (c) Zoomed-in view of protein-ligand binding pockets, with yellow dashed lines indicating hydrogen bonds between ligand and protein, and purple dashed lines representing salt bridges. CODE-designed binders form structurally compact and energetically favorable pockets with their targets. (d) Distribution of pairwise TMscores between CODE-optimized binders (green)/original BoltzDesign1 binders (purple) and wild-type proteins. (e) Bar plots compare AF3 success rates between CODE-optimized and original BoltzDesign1 approaches under two filtering criteria. To evaluate intrinsic design capabilities, BoltzDesign1 output structures were used directly without LigandMPNN redesign. (f) Box plots showing secondary structure distribution in CODE-optimized binders. From left to right: percentages of helix, loop/coil, and beta sheet content. Default BoltzDesign1 helix loss weight parameters were applied (minimum -0.3, maximum 0).The target molecules SAM, IAI, FAD, and OQO correspond to the wild-type PDB structures 7C7M, 5SDV, 7BKC,  and 7V11, respectively.}
    \label{binder}
\end{figure}

The design of protein binders that can specifically interact with molecular targets to mediate catalytic activity, molecular recognition, structural support, and protein-biomolecule interactions represents a critical goal in synthetic biology and therapeutic development \cite{yang2023alphafold2,watkins2015structure,lu2020recent}. Deep learning-based computational methods have made great progress in recent years with the improving  protein design success rate. One popular method involves fine-tuning structure prediction models into diffusion models to generate novel structures, such as RfDiffusion \cite{watson2023novo} and RfDiffusionAA \cite{krishna2024generalized}. Another method does not require additional training and directly uses the structure prediction model back propagation to iteratively optimize the sequence until convergence. The back-propagation method can model sequence and structure simultaneously, and has the unique ability to predict the structure of ligands and binders during the design process, thereby being able to explore the induced fit effect during binding, which is currently unavailable to other computational methods \cite{cho2025boltzdesign1}. More importantly, this method allows for custom loss function optimization without retraining the model, which provides the possibility for flexible programmability of binder design. Naturally, since the CODE we proposed reflects topological frustration to a certain extent, it can provide strong guidance information for the topological conformation design near the binding site, so it is very suitable for improving the quality of binder design.

To illustrate that CODE can be used as a guide for binder design, we follow BoltzDesign1 \cite{cho2025boltzdesign1}, which implements iterative sequence design by inverting the all-atom prediction model Boltz-1 \cite{wohlwend2024boltz}, and directly optimizes the probability distribution of pairing features through the predicted distribution graph. Specifically, BoltzDesign1 combines the confidence module with the Pairformer to achieve joint optimization based on distance map loss and confidence scores. The confidence module represents the fit between structures sampled from the diffusion module and the pairwise feature probability distributions. In certain scenarios, such as protein-DNA/RNA interactions where pairwise features are ambiguous, additional sampling within the diffusion module may better assess the accessible structural space. To enable loss backpropagation from the confidence module to the inputs, BoltzDesign1 modified the process to allow gradient flow between the Pairformer and confidence module. Addtionally, we calculate CODE when calling the structure diffusion module as an additional loss function to guide binder design compared with BoltzDesign1. As a simple verification, we follow the setting in BoltzDesign1: use the small molecules tested in RfDiffusionAA as a benchmark to design protein binders. The workflow and calculation steps are as follows:

\begin{equation}
\hat{s}_{trunk} , \hat{z}_{trunk} = \text{Pairformer}({s_{trunk}}, {z_{trunk}}, S_{\text{input}})
\end{equation}

\begin{equation}
xyz_{atom} = \text{Structure Module}(\hat{s}_{trunk}, \hat{z}_{trunk}).{stop\_gradient}
\end{equation}

\begin{equation}
\text{Confidence score} =\text {Confidence Module}(\hat{s}_{trunk}, \hat{z}_{trunk}, xyz_{atom}, S_{\text{input}})
\end{equation}
where $s_{trunk}$ represents the single-sequence representation, $z_{trunk}$ denotes the pairwise representation, and $S_{input}$ is the initial sequence input. $xyz_{atom}$ corresponds to the denoised 3D structure coordinates. Here, $\hat{s}_{trunk}$ and $\hat{z}_{trunk}$ are the outputs from the Pairformer, where gradients are allowed to flow through to the confidence module.

\paragraph{CODE guides BoltzDesign1 to achieve higher in silico success rates}
After performing hallucination design through Boltz1, BoltzDesign1 also calls LigandMPNN \cite{dauparas2025atomic} to redesign the preliminary designed sequence. However, we only want to explore the basic design capabilities without considering the unfair information introduced by LigandMPNN. So we disabled all post-sequence design. BoltzDesign1 presented in silico success rates of binder design for four small molecules: IAI, FAD, SAM and OQO (Figure \ref{binder}a). For each ligand, we used the default settings of BoltzDesign1 and the configuration with CODE loss to design 30 structures. Then Alphafold3 \cite{abramson2024accurate} was used to predict the complex structure. We evaluate these results based on structure confidence scores, specifically the pLDDT from AlphaFold3 for the entire complex and the ipAE for the binding interface confidence between the binder and the small molecule. The calculation of the success rate is divided into two modes. In strict mode, complex $pLDDT >70$ and $ipAE <10$ are required, and in relaxed mode, complex $pLDDT>70$ and $ipAE<15$ are required \cite{bennett2023improving,harteveld2024exploring,jendrusch2025alphadesign}.

As shown in Figure \ref{binder}b, CODE-based design binders have strong structural consistency between models. For four targets, these designed binders optimized based on CODE loss well show the interactions between proteins and ligands, including hydrogen bonds, hydrophobic interactions and $\pi-\pi$ interactions (Figure \ref{binder}c). On the first three targets, the design samples with CODE loss restrictions all showed better AF3 success rates than the original BoltzDesign1. Specifically, on SAM, we achieved a strict success rate of 67\%, a relative improvement of 11.6\% compared to BoltzDesign1's 60\%. On IAI, we reached a strict success rate of 73\%, a 21.6\% relative improvement over BoltzDesign1's 60\%. On FAD, we attained a strict success rate of 65\%, a 14\% relative improvement compared to BoltzDesign1's 57\%. In terms of relaxed success rate, our model performed on par with the baseline model, with a 5\% relative improvement on IAI. On OQO, due to the clearer binding mode, both methods achieved good comparable results. 

These significant improvements demonstrate that the topological frustration information introduced by CODE effectively captures interactions between biological molecules, providing novel insights for binder design that BoltzDesign1 overlooks by ignoring key structural topological constraints.

\paragraph{CODE preserves the diversity of generated samples}
To evaluate whether the introduction of topological frustration loss would reduce the diversity of generated structures, we compared with the original Boltzdesign1. For each target molecule, we calculated the TMscore of the pairwise generated binder structure and the wild type, without any post-processing filtering. We show the distribution of TMscore in Figure \ref{binder}d. In general, the introduction of topological frustration CODE will emphasize the local structure maintained by the binder design, but will not harm the diversity of generated structures, and even achieve better diversity for some targets. For the targets 5SDV and 7BKC, the average TMscore of the binder designed with CODE loss is 0.199 and 0.110, respectively, while the TMscore without CODE loss is 0.224 and 0.115. The addition of CODE loss enhances diversity in designs for these two targets, outperforming BoltzDesign1, which produces more similar samples. For targets 7C7M and 7V11, although the average TMscore of the binder designed with CODE loss is 0.232 and 0.212, respectively, and the TMscore without CODE loss is 0,226 and 0.202, they are almost consistent, with no obvious changes in diversity.

We also evaluated the distribution of secondary structure in the designed binders in Figure \ref{binder}f. Among the four binder designs, an increase in the helix proportion was found to be a favorable design factor. In the binder designs for SAM, IAI, and FAD, we observed a higher percentage of helix while reducing the content of the loop region. In contrast, in OQO, the trend was the opposite, and this result showed consistency with the design success rate. No clear pattern was observed in the sheet distribution; however, increasing the percentage of sheet could stabilize the overall structure of the binder. In the future, a corresponding loss term could be introduced to enhance its content.

Topological frustration refers to energetically conflicted regions in protein folding that exhibit high conformational flexibility. In binder design, targeting these dynamic sites on protein surfaces can enhance induced fit or conformational selection mechanisms. Due to their functional significance, topologically frustrated regions are often ideal drug targets. By leveraging local frustration energy, CODE optimizes binder design to improve affinity for target proteins.


\subsubsection{Enzyme Catalytic Active Sites Identification}

\begin{figure}[htbp]
    \centering
\includegraphics[width=1\textwidth]{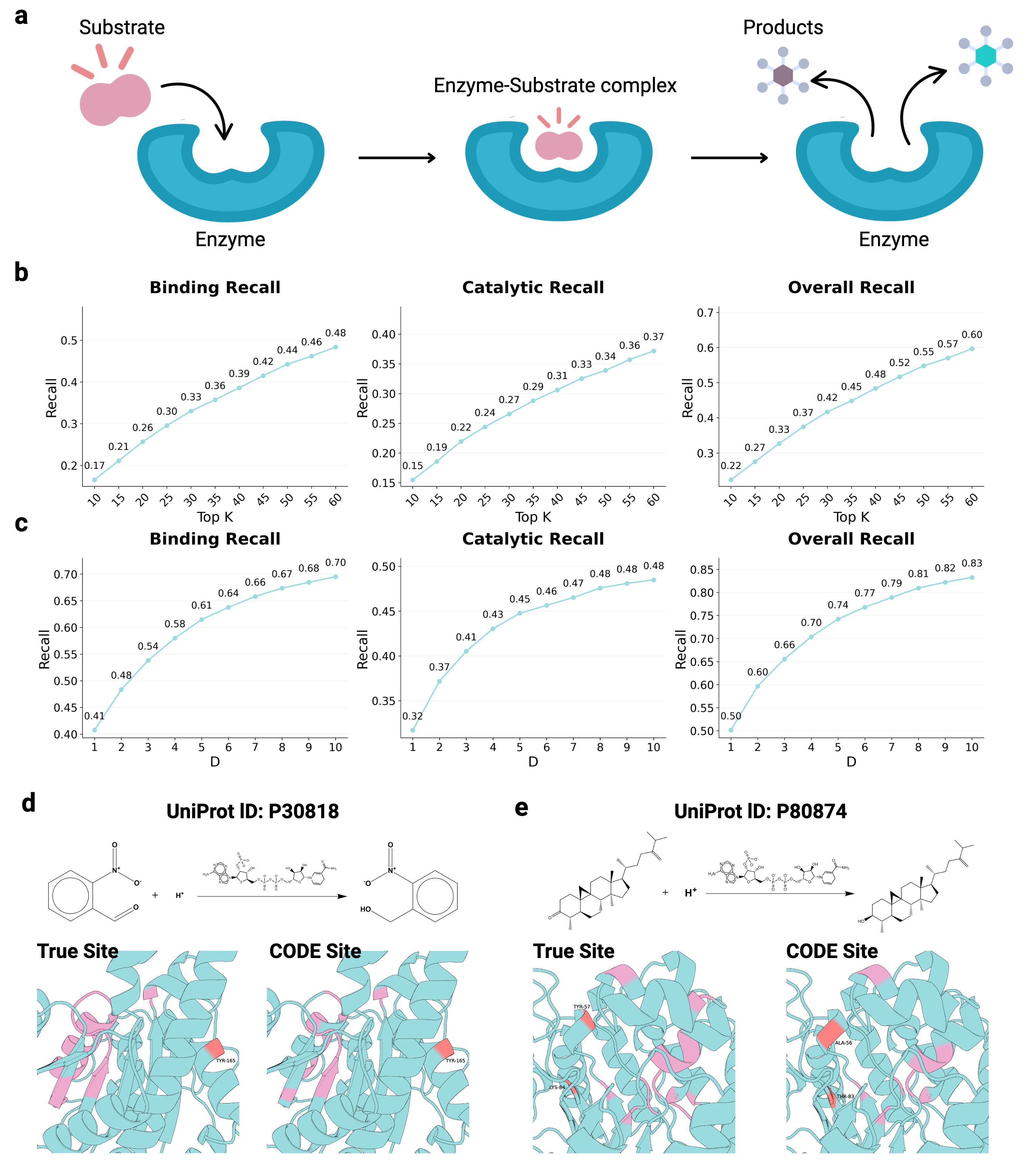}
    \caption{\textbf{Residue level CODE ranking can reveal the catalytic active site and binding site of the enzyme.} (a) Schematic diagram of the process of enzyme catalyzing substrate reaction. (b) Under the condition $D = 2$, the CODE values are sorted in ascending order, with the top $K$ selected as recognition sites. The recall rates for binding sites, catalytic sites, and overall sites are calculated and presented. (c) After selecting the top 60 values and sorting them in ascending order as recognition sites, recall rates for binding sites, catalytic sites, and overall sites are calculated across varying tolerance values $D$. (d-e) Chemical reaction equations catalyzed by enzymes in two case studies (UniProt ID: P30818 and P80874) are presented. The actual enzyme catalytic sites (highlighted in orange), binding sites (highlighted in pink), and CODE-based sorting recognition results are compared. CODE demonstrates a significant recall rate, serving as an effective tool for identifying key enzyme sites.}
    \label{enzyme}
\end{figure}

Accurate identification of enzyme active sites remains a fundamental challenge in biochemistry, despite its critical importance for understanding metabolic processes and advancing drug development. While DNA sequencing has generated millions of enzyme sequences, fewer than $0.7\%$ have properly annotated active sites due to the complexity of predicting these functional regions \cite{uniprot2023uniprot}. This bottleneck stems from the need to understand specific enzyme-substrate interactions, reaction mechanisms and the ability to distinguish between different site types. The scarcity of high-quality training data makes direct training of predictive methods particularly challenging \cite{petrova2006prediction,jones1999protein,teukam2024language,wang2024multi}. However, CODE uses topological frustration theory in protein folding to gain universal structural insights, providing a practical solution for data-scarce scenarios without requiring specific training.

We implemented the residue-level CODE algorithm to prioritize amino acid rankings. The rationale is rooted in evolutionary biophysics: residues critical for function experience strong purifying selection, mutate infrequently, and therefore are more likely to occupy structurally stable regions with minimal topological frustration \cite{todd2002plasticity}. CODE is expected to reach its minimum values at these functionally critical locations.

Specifically, we adopted the test set released by EasIFA \cite{wang2024multi}, comprising 853 non-redundant enzymes with residue-level annotations for ligand-binding sites, catalytic sites and others. For every enzyme, we computed CODE score $s_{i}$ for residue $i$ and ranked residues in ascending order. CODE was applied in a fully zero-shot manner without fitting model parameters to the dataset. We introduced the Relaxed Overlap Rate (ROR), a novel metric to quantify CODE’s detection accuracy by comparing ranked predictions to curated functional annotations with adjustable positional tolerance. Let $\mathcal{A}$ denote the intervals of annotated functional residues, $\mathcal{P}$ the intervals covered by the top-$k$ CODE-ranked residues, and $D$ the allowed sequence-index tolerance. Then
\begin{equation}
\mathcal{P}_{D} = \left\{\, i \,\middle|\, \exists\, j \in \mathcal{P},\; |i-j| \le D \right\}
\label{eq:pd}
\end{equation}

\begin{equation}
\mathrm{ROR} = \frac{\lvert \mathcal{A} \cap \mathcal{P}_{D} \rvert}{\lvert \mathcal{A} \rvert}
\label{eq:ror}
\end{equation}

We first analyzed the impact of different Top K selections on identification Recall, which is shown in Figure \ref{enzyme}b. We observed that as K increased, Recall continuously improved, but the rate of increase gradually slowed. In Top 60, we achieved a Binding Site Recall of 48\%, a Catalytic Site Recall of 37\%, and an overall Recall of 60\%. Given that the amino acid length of these enzymes is generally above 500, achieving a 60\% Recall by unsupervised selection of 10\% of the sites without additional training is a highly promising result. Then we fixed the Top 60 selection and observed the changes in Recall with varying tolerance levels, which is shown in Figure \ref{enzyme}c. We noted that as the tolerance (D) increased, Recall continuously improved, but the rate of increase gradually slowed. At D=10, we achieved a Binding Site Recall of 70\%, a Catalytic Site Recall of 48\%, and an overall Recall of 83\%. Since CODE is not an algorithm specifically designed for enzyme site identification and topological frustration is influenced by surrounding amino acids, CODE achieves  high identification accuracy within a reasonable error tolerance range. Finally, we selected two representative enzymes (UniProt ID: P30818 and P80874) in Figure \ref{enzyme}d and \ref{enzyme}e for detailed, visually intuitive case studies.

Catalytic residues are often responsible for proton relay, nucleophilic attack, or transition-state stabilization, but the chemistry can proceed only if these side chains are held in a geometrically precise and electrostatically pre-organized constellation. Evolution therefore embeds them in a highly cooperative network of contacts that dampens local strain and maximizes structural rigidity. In the CODE formalism, such cooperativity is reflected by low CODE scores, i.e., minimal topological frustration. Strikingly, we observe similarly depressed CODE values for ligand-binding pockets located adjacent to the catalytic center. This convergence toward regions of low frustration provides a coherent biophysical rationale for the frequent spatial overlap of binding and catalytic motifs. Because CODE requires only a structure prediction model and no task-specific training, it offers a scalable route to annotate the millions of orphan enzyme sequences currently lacking functional labels. By coupling CODE-based priors with sparse experimental data, one can envisage active-site mapping at proteome scale, accelerating enzyme engineering and structure-guided drug discovery.

\subsubsection{Drug-Resistant Mutants Prediction}

\begin{figure}[htbp]
    \centering
\includegraphics[width=1\textwidth]{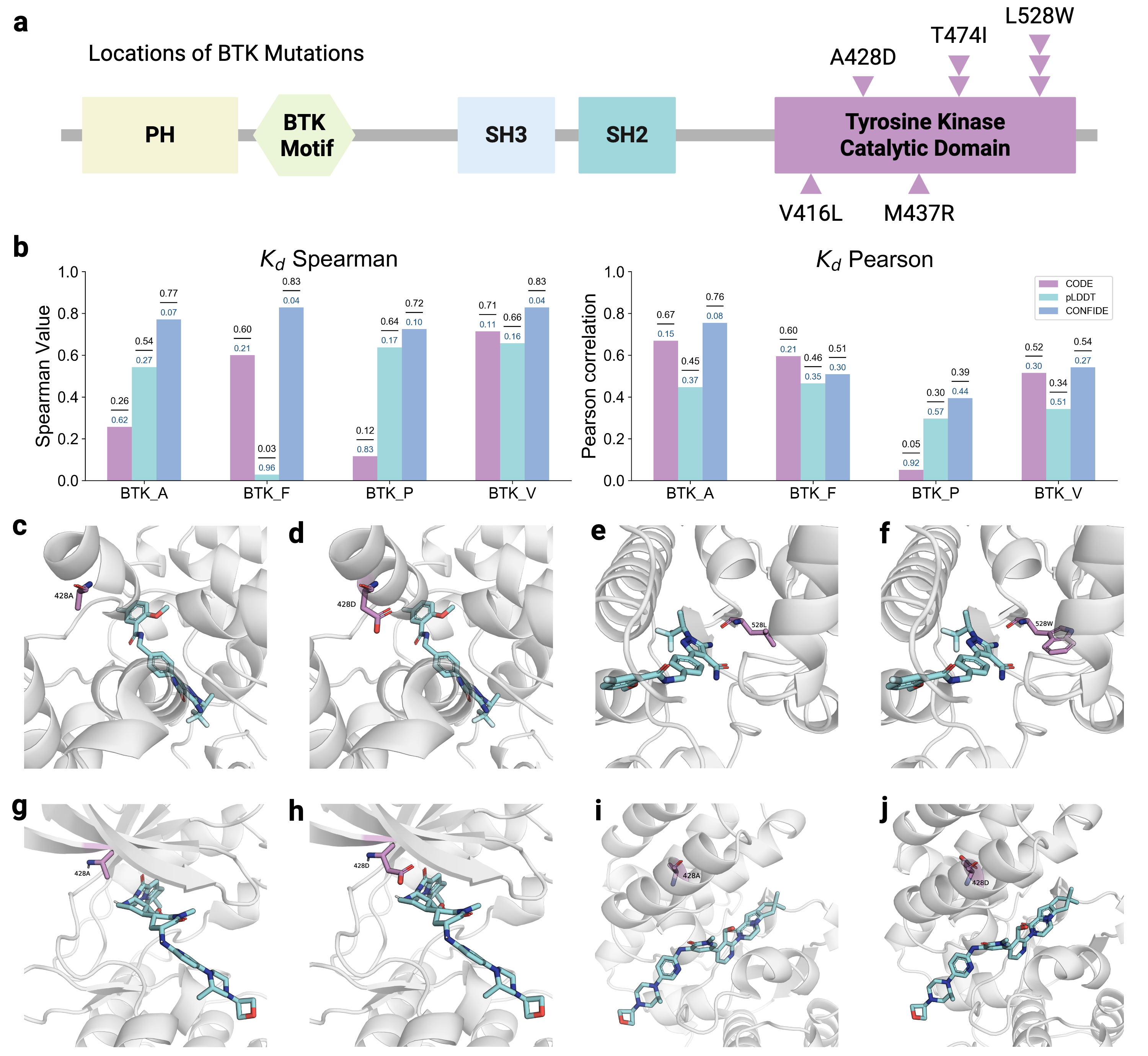}
    \caption{\textbf{CODE exhibits high effectiveness in predicting mutation-induced drug resistance effects.} We focus on the relevance of five BTK protein mutations to changes in affinity (expressed as $K_d$) for four non-covalent inhibitors (Pirtobrutinib as BTK\_P, ARQ-531 as BTK\_A, Vecabrutinib as BTK\_V and Fenebrutinib as BTK\_F). (a) shows BTK mutations outside the BTK C481 residue 
    found in patients with resistance to pirtobrutinib. Each 
    individual occurrence of a mutation is depicted as an arrowhead. PH denotes pleckstrin homology domain, and SH Src homology domain.
    (b) Spearman and Pearson correlation coefficients are presented for CODE, pLDDT, and CONFIDE in relation to $K_d$, with each bar displaying the correlation coefficient value (in black) and the p-value (in blue), separated by a hyphen. The predicted structures that conform to the real mechanism when there is no hallucination are shown in (c-f). (c)(e) show the reasonable interaction between the wild-type protein and Pirtobrutinib and appropriate pocket size in 428A and 528L. (d) shows A428D mutation which introduces a side chain that significantly blocks the kinase pocket entrance, causing substantial steric hindrance with the drug molecule and eliminating binding affinity. (f) shows L528W mutation which increases steric hindrance but causes less severe conflict than A428D, allowing weak binding affinity at 100 $\mu M$. Figures (g-j) illustrate that pLDDT fails to accurately reflect structural interpretations of affinity changes in the presence of hallucinated predictions. (g) shows the reasonable interaction between the wild-type protein and Fenebrutinib and appropriate pocket. (h) shows significant increase in steric hindrance for the A428D mutation structure obtained through homology modeling, which is consistent with the experimental observation that no binding has been detected. (i) shows predicted structure by Boltz1 between the wild-type protein and Fenebrutinib. (j) shows the predicted A428D mutation structure by Boltz1, exhibiting no significant steric hindrance, which fails to explain the lack of binding and results in a very low correlation between pLDDT and $K_d$.}
    \label{btk}
\end{figure}


Drug resistance represents one of the most formidable obstacles in modern cancer therapeutics, particularly affecting the clinical efficacy of targeted kinase inhibitors that have transformed treatment paradigms for numerous malignancies. Bruton's tyrosine kinase (BTK) exemplifies this challenge, serving as a critical therapeutic target in B-cell malignancies including chronic lymphocytic leukemia and mantle cell lymphoma \cite{wang2022mechanisms}. While BTK inhibitors such as pirtobrutinib have demonstrated remarkable clinical success, the inevitable emergence of resistance mutations necessitates sophisticated approaches to predict and overcome therapeutic resistance before it manifests clinically \cite{joseph2024impact}.

The molecular basis of kinase inhibitor resistance typically involves structural perturbations that alter the binding landscape of the ATP-binding pocket or allosteric regulatory sites. These mutations can range from subtle amino acid substitutions that modestly reduce inhibitor affinity to dramatic structural rearrangements that completely abrogate drug binding. Understanding how specific mutations translate to quantitative changes in binding affinity is crucial for developing next-generation inhibitors and optimizing treatment strategies. However, experimental characterization of all possible resistance mutations is impractical, highlighting the urgent need for computational approaches that can accurately predict the impact of mutations on inhibitor binding.

We evaluated the performance of CODE and CONFIDE in predicting drug resistance effects using well-characterized BTK mutations known to reduce the efficacy of pirtobrutinib, a clinically approved non-covalent BTK inhibitor. Our analysis extended beyond pirtobrutinib to include three additional non-covalent BTK inhibitors: ARQ-531, Vecabrutinib and Fenebrutinib, allowing us to examine whether CODE could accurately capture resistance patterns across multiple compounds targeting the same kinase. Binding affinities between various compounds and both wild-type and mutant forms of purified BTK protein were quantified through surface plasmon resonance measurements, yielding dissociation equilibrium constants (K\textsubscript{D}). These experimental data were sourced from earlier research \cite{wang2022mechanisms}. We inputted the wild-type or mutant protein sequences along with ligands into Boltz-1 to generate predicted structures, from which we extracted confidence scores and combined them with topological frustration values (CODE) to calculate CONFIDE scores. Finally, we computed spearman and pearson correlation coefficients between confidence scores, CODE, and CONFIDE against  K\textsubscript{D} values. In this task, since we are more concerned about the ranking ability of the scores, we mainly consider the performance of the spearman correlation coefficient.


The correlation between different scores and $K_d$ is shown in Figure \ref{btk}b. For drugs Vecabrutinib and Fenebrutinib, CODE takes the lead, plddt provides additional information, and CONFIDE is the most comprehensive. For drug Vecabrutinib, the absolute Spearman correlation of pLDDT was 0.66, while CODE improved this to 0.71 (a 7.6\% relative increase over pLDDT). CONFIDE achieved an absolute correlation of 0.89, which represents a 25.4\% relative improvement over CODE and a 34.8\%  relative increase compared to pLDDT. Similarly, for drug Fenebrutinib, pLDDT achieved the lowest absolute correlation at 0.03, while CODE increased this to 0.60 (a 1900\% relative improvement). CONFIDE again performed best, achieving an absolute correlation of 0.83, which is 38.3\% higher than CODE and 2758\% better than pLDDT. 

However, for drugs Pirtobrutinib and ARQ-531, the trend of pLDDT and CODE reversed. Specifically, for drug Pirtobrutinib, CODE achieved an absolute spearman correlation of 0.12, pLDDT improved this to 0.64, while CONFIDE reached 0.73, representing a further relative improvement of 14\% over pLDDT and a total relative improvement of 508\% compared to CODE. Similarly, for drug ARQ-531, CODE showed an absolute correlation of 0.26, while pLDDT increased to 0.54. CONFIDE again outperformed both, achieving 0.77, which is 42.6\%  relative higher than pLDDT and an overall relative 196\% increase compared to CODE. These results demonstrate that CONFIDE consistently integrates the strengths of both topological and energy frustration, yielding the highest predictive accuracy for all these four drugs.

In addition to its strong sorting ability, CONFIDE also basically maintains the leading Pearson correlation that reflects the linear correlation. In drugs ARQ-531, Fenebrutinib, and Vecabrutinib, CODE takes the lead, and pLDDT serves as auxiliary information. In drug ARQ-531, CODE's spearman correlation reached 0.67, pLDDT was 0.45, and CONFIDE surpassed both to reach 0.76, achieving a relative improvement of 13.4\% and 68.9\%, respectively. In drug Fenebrutinib, CODE had the best spearman correlation, reaching 0.60, pLDDT was 0.47, and CONFIDE was 0.51. Because of the consideration of stronger sorting ability, CONFIDE made certain sacrifices in linear relationships. In drug Vecabrutinib, CODE's spearman correlation reached 0.52, pLDDT was 0.34, and CONFIDE surpassed both to reach 0.54, achieving a relative improvement of 58.8\% compared to pLDDT. However, in drug Pirtobrutinib, the Pearson correlation coefficients of pLDDT and CODE are 0.30 and 0.05, respectively. CONFIDE is still able to integrate the information of the two to obtain the best result of 0.39, achieving a relative improvement of 33.3\% and 700\%.

These results suggest that while CONFIDE consistently outperforms the other methods, the relative contributions of topological and energy frustration vary depending on the drug. For drugs Pirtobrutinib and ARQ-531, energy frustration plays a more significant role, as reflected in the greater improvement of pLDDT over CODE. In contrast, for drugs Vecabrutinib and Fenebrutinib, topological frustration appears to dominate, as CODE outperforms pLDDT and provides a solid foundation for CONFIDE's superior performance. This highlights the context-dependent nature of these methods and emphasizes the value of combining both metrics in CONFIDE for robust predictive power across diverse drug classes.

\paragraph{In-depth Case Studies}
We conduct a more in-depth case analysis based on the structural biology perspective of steric hindrance. We will analyze the two situations of whether hallucination occurs or not: in the absence of hallucinations, we investigated the fundamental mechanism underlying pLDDT’s functionality. When hallucination occurs, CODE provides additional information to help understand the predicted structure more comprehensively, so as to avoid being misled by hallucination. According to previous studies\cite{wang2022mechanisms,wong2024mutation,mato2023pirtobrutinib}, drug-resistant mutations mainly play a role through steric hindrance - the L528W mutation inside the binding pocket and the A428D mutation at the entrance will introduce a large side chain to produce spatial constraints, thereby physically hindering the optimal positioning of the inhibitor.

In Figure \ref{btk}c-f, we show the predicted structure that conforms to the real mechanism when there is no hallucination. Figure \ref{btk}c shows the interaction between the wild-type protein and Pirtobrutinib. Due to the reasonable ligand conformation and appropriate pocket size, the wild type has a strong affinity. However, when the A428D mutation occurs, that is, the structure shown in Figure \ref{btk}d, the side chain will significantly block the entrance of the kinase pocket, resulting in a large steric hindrance between the protein and the drug molecule, so no binding affinity is observed. Similarly, Figure \ref{btk}e shows the interaction between the wild-type protein and Pirtobrutinib, and Figure \ref{btk}f shows the L528W mutation. Although the L528W mutation also increases steric hindrance, it does not cause such a severe conflict as A428D. Therefore, L528W can also be observed to bind weakly at 100 $\mu M$. The reason why plddt can show a relative strong Pearson correlation of 0.64 under this drug is that the predicted structure shown in Figures a-d can correctly reflect the above mechanism.

However, when hallucination occurs, plddt cannot be served as a good judgment standard. Figure \ref{btk}g shows the interaction between the wild-type protein and the Fenebrutinib drug. We used homology modeling to obtain the A428D mutation structure in Figure \ref{btk}h. In reality, this mutation will produce large steric hindrance, which is consistent with the experimental observation that no binding has been detected. However, Figures \ref{btk}i-j show the protein-drug binding structure before and after the mutation predicted by Boltz1. As can be seen, no significant increase in steric hindrance is observed in the predicted structure, so when plddt evaluates the structure, it is likely that it will not recognize that this is an unreasonable conformation, thus giving a misleadingly high confidence. However, CODE can effectively capture such spatial steric effects, accurately reflecting binding affinity and aligning closely with experimental results.

CODE reflects topological frustration, providing valuable mechanistic insights for comprehensive analysis, complementing the structural predictions observed in our BTK resistance studies. Higher topological frustration values may indicate a more severe loss of binding ability, while a moderate decrease in frustration values may correspond to a moderate increase in affinity. Regions with minimal topological frustration in wild-type BTK represent evolutionarily conserved structural elements that are critical for inhibitor recognition, explaining why mutations such as C481S within these low-frustration regions are able to maintain high binding affinity by retaining essential protein-ligand contacts. The differential correlation patterns observed between different inhibitors - Vecabrutinib and Fenbutinib show clear structure-activity relationships, while Pirotinib and ARQ-531 do not - may reflect different reliance on specific frustration landscapes. Most importantly, the case of A428D on Fenebrutinib, which exhibits a high confidence score despite a significant loss in experimental affinity, highlights how topological frustration analysis can capture subtle local destabilizing effects that are not detectable from static structural predictions alone, leading to a more nuanced understanding of how mutations may disrupt binding through mechanisms of  steric interference.

\subsubsection{Inhibitor Affinity Prediction}

\begin{figure}[htbp]
    \centering
\includegraphics[width=1\textwidth]{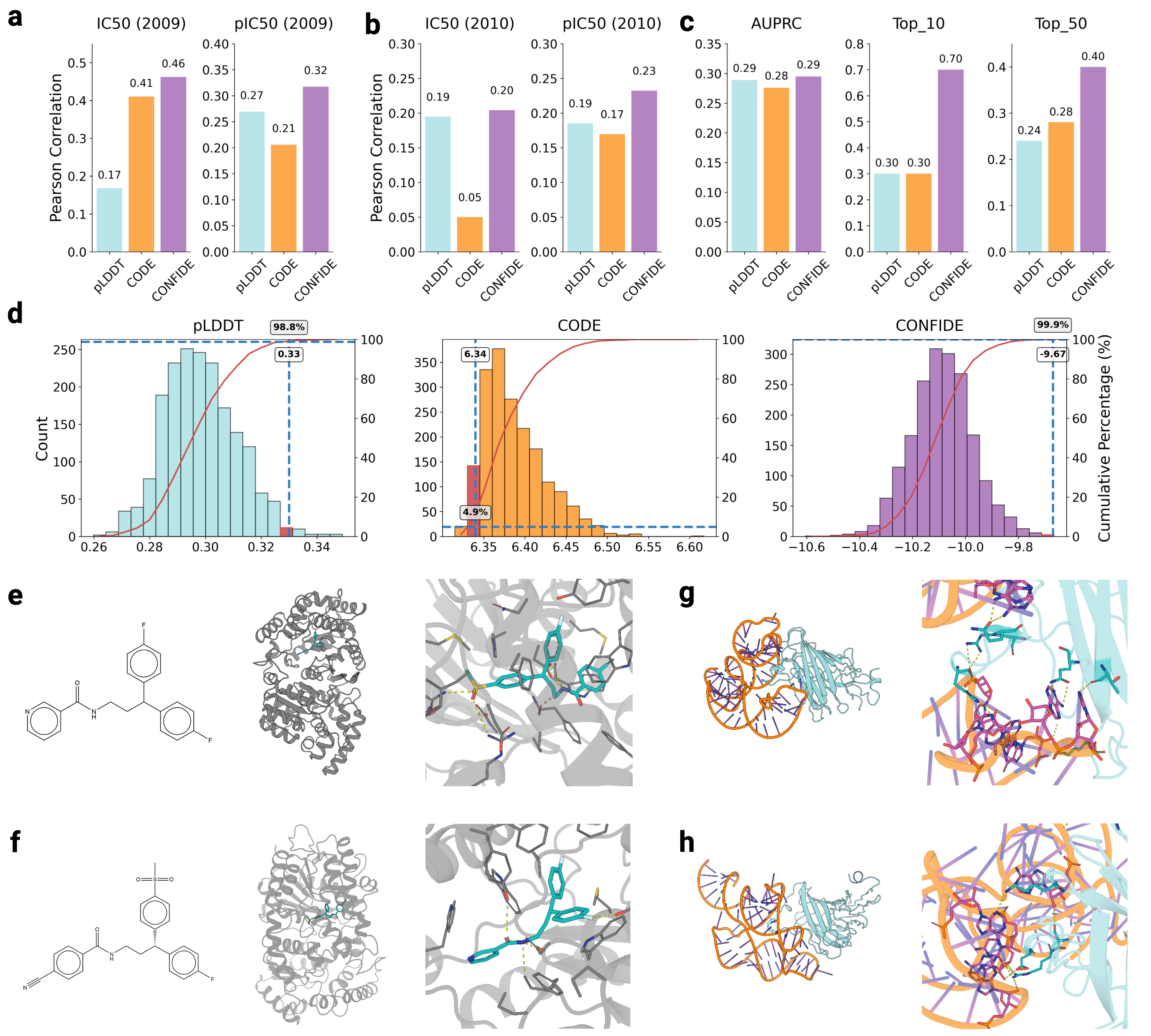}
    \caption{\textbf{CODE is an unsupervised virtual screening tool.} (a-b) show the Pearson correlation of CODE, pLDDT, and CONFIDE with activity data (\(\text{IC}_{50}\) and \(\text{pIC}_{50}\)) for the sEH2009 and sEH2010 datasets, respectively. (c) presents the AUPRC classification metrics for CODE, pLDDT, and CONFIDE in RNA aptamer screening, alongside the hit rates for the top 10 and top 50 candidates. (d) illustrates the distribution of pLDDT, CODE, and CONFIDE scores on the RNA aptamer dataset, with the red highlighted region indicating the positions of these scores corresponding to a top 10 hit aptamer in the distribution. (e-f) show the binding interfaces of two screened sEH inhibitors with their corresponding molecules and proteins. (g-h) show the binding interfaces of two screened RNA aptamers. }
    \label{screening}
\end{figure}

Soluble epoxide hydrolase (sEH) represents a compelling therapeutic target for cardiovascular, inflammatory, and neurological disorders due to its central role in regulating bioactive lipid mediators \cite{rose20101,eldrup2009structure}. This enzyme catalyzes the hydrolysis of epoxyeicosatrienoic acids (EETs), which possess anti-inflammatory, vasodilatory, and cardioprotective properties. By inhibiting sEH, these beneficial epoxide metabolites are preserved, offering a therapeutic strategy for conditions ranging from hypertension to chronic pain. The development of potent and selective sEH inhibitors has therefore attracted significant attention in medicinal chemistry, necessitating sophisticated approaches to identify compounds with optimal binding characteristics \cite{shen2012discovery}.

The challenge of inhibitor screening and optimization for sEH exemplifies a broader problem in structure-based drug design: predicting how subtle molecular modifications translate to significant changes in binding affinity. Traditional high-throughput screening approaches, while comprehensive, are resource-intensive and may miss nuanced structure-activity relationships that emerge from the dynamic nature of protein-ligand interactions. This limitation underscores the need for computational methods that can effectively discriminate between compounds with varying affinities within related chemical series.

We inputted all protein-ligand pairs into Boltz-1 for structure prediction and obtained the confidence scores as energy frustration metrics. By combining the confidence score and the topological frustration-aware CODE, we calculate CONFIDE. Finally, we computed Pearson and Spearman correlation coefficients between the CODE, CONFIDE and confidence scores and both IC50 and pIC50 values. The experimental results are shown in Figure \ref{screening}.

Our correlation analysis of the two datasets shows that CONFIDE's Pearson correlation consistently outperforms the pLDDT and CODE metrics alone. In the 2009 dataset, CODE showed a stronger Pearson correlation on IC50 (0.41 vs. 0.17 for pLDDT, a relative improvement of 142\%), while CONFIDE surpassed both to 0.46, a relative improvement of 170\% over pLDDT. On pIC50, pLDDT showed a better Pearson correlation (0.27 vs. 0.21 for CODE), while CONFIDE surpassed both to 0.32, an improvement of 18.5\% over pLDDT. This pattern changed in the 2010 dataset. For IC50, pLDDT has a higher Pearson correlation coefficient (0.195 vs. CODE's 0.050), and CONFIDE still maintains the lead at 0.204, a relative improvement of 4.6\% compared to pLDDT. For pIC50, pLDDT also has a higher Pearson correlation coefficient than CODE (0.185 vs. CODE's 0.170). CONFIDE reached 0.232, a 25.4\% improvement over the second highest pLDDT. These results clearly show that CONFIDE is the most superior metric, significantly improving the performance of single energy or topological metrics in all comparisons, and also highlighting the complementary advantages of pLDDT and CODE when prioritizing linear relationships.

CODE offers a novel understanding of binding affinity prediction. In the context of protein-ligand complexes, regions of low topological frustration typically correspond to structurally conserved binding sites that have evolved to accommodate specific molecular interactions with high fidelity. These minimally frustrated regions often correlate with critical binding determinants where even minor structural perturbations can significantly impact affinity. Conversely, areas of higher frustration may represent more adaptable binding environments that can tolerate diverse ligand modifications without dramatic affinity changes. By mapping topological frustration patterns within sEH-inhibitor complexes, we can potentially identify which molecular features are most crucial for maintaining high-affinity interactions, thereby enhancing our ability to predict and rank compound affinities based on their interactions with these frustration-sensitive regions.

\subsubsection{Virtual Screening of RNA Aptamers}

RNA aptamers represent an important class of therapeutic and diagnostic molecules that combine the specificity of antibodies with the versatility and synthetic accessibility of nucleic acids \cite{brody2000aptamers,coutinho2019rna}. These short, structured RNA sequences fold into unique three-dimensional conformations that enable highly specific binding to target proteins, cells, or other biomolecules. Green fluorescent protein (GFP) serves as an excellent model target for aptamer development due to its well-characterized structure, widespread use as a research tool, and potential applications in biosensing and cellular imaging. The identification of high-affinity GFP-binding aptamers from large candidate pools represents both a significant technical challenge \cite{morsch2023aptabert,chandola2020aptamers}.

Traditional experimental aptamer selection through SELEX (Systematic Evolution of Ligands by EXponential enrichment) is inherently time-consuming and resource-intensive, often requiring multiple rounds of selection and amplification to identify optimal binders from libraries containing up to $10^{15}$ unique sequences \cite{ellington1990vitro,stoltenburg2007selex}. Virtual screening approaches offer a valuable alternative, enabling the computational evaluation of vast aptamer libraries to prioritize candidates for experimental validation \cite{huang2024protein,nori2024rnaflow}. However, the success of such approaches critically depends on accurate prediction of binding affinities and the ability to distinguish high-affinity aptamers from weak or non-specific binders within structurally diverse RNA populations.

We sought to identify high-affinity GFP-binding aptamers from an extensive candidate library. Our analysis utilized a curated dataset of GFPapt mutants by prior work \cite{huang2024protein}, originally derived from the earlier study \cite{shui2012rna}. The dataset encompasses aptamers exhibiting a broad spectrum of binding affinities, with dissociation constants (K\textsubscript{d}) ranging from 0 nM to 125 nM. For classification purposes, aptamers demonstrating K\textsubscript{d} values below 10 nM were designated as high-affinity binders and classified as positive hits for subsequent analysis. We inputted all protein-RNA pairs into Boltz{-}1 for structure prediction, from which we extracted confidence scores and combined them with topological frustration values (CODE) to calculate CONFIDE scores. Finally, we used these three scores to compute three classification metrics: AUPRC, Precision@10, and Precision@50.

In Figure \ref{screening}c, we reported the results for the three scores. We observed that CONFIDE demonstrates a significant advantage in screening for the highest-affinity aptamers. For the top 10 screening, CONFIDE achieved a hit rate of 70\%, a 133\% improvement compared to pLDDT's 30\%. For the top 50 screening, CONFIDE reached a hit rate of 40\%, surpassing pLDDT's 24\% and CODE's 28\%, representing a 66.6\% improvement over pLDDT. To further analyze the properties of the screened candidate aptamers, we examined the distribution of pLDDT, CODE, and CONFIDE scores in the candidate library, as shown in Figure \ref{screening}d. We highlighted representative aptamers in the top 10 with red, noting that their pLDDT values were distributed in the higher range, while CODE scores were lower, ultimately leading to higher CONFIDE score rankings, resulting in their selection.

Regions of minimal topological frustration in GFP likely correspond to evolutionarily conserved structural elements that form stable, high-fidelity binding interfaces. Aptamers that interact primarily with these low-frustration regions may exhibit enhanced binding stability and specificity, as these areas represent energetically favorable interaction sites that are less tolerant of suboptimal contacts. Conversely, aptamer binding modes that engage highly frustrated regions of GFP may result in weaker or more promiscuous interactions. By incorporating topological frustration mapping into virtual screening workflows, we can potentially enhance our ability to identify aptamers that form thermodynamically stable complexes with GFP, while simultaneously gaining mechanistic insights into the structural determinants of RNA-protein recognition that could inform rational aptamer design strategies.




\section{Methods}\label{method}
\subsection{Overview}
The current AlphaFold3 confidence metrics, especially pLDDT, often give the hallucination of high confidence but low prediction accuracy when faced with many difficult structure prediction scenarios. To solve this problem, we proposed CODE and COFIDE. Extensive research suggests that protein folding occurs in a complex energy landscape and is mainly affected by two different types of frustration: energy frustration and topological frustration \cite{brockwell2000protein}. Although evolutionary selection minimizes the energy frustration of natural proteins, topological frustration remains another important intrinsic property that determines folding pathways and dynamics. CODE extracts the embedding of each layer in the diffusion transformer of the diffusion module in alphafold3, and analyzes the changing trajectory of the embedding in high-dimensional space to reflect the model's own estimation of topological frustration information. CONFIDE is built on the basis of both CODE and pLDDT (or confidence score), providing the most comprehensive modeling perspective on the folding landscape. In the following sections, we will introduce each score one by one and their contribution to understanding the quality assessment of predicted structures.

\subsection{Chain of Diffusion Embedding}
To characterize topological frustration in AlphaFold3, we draw inspiration from the Chain-of-Embedding (COE) paradigm developed for transformer-based language models \cite{wang2024latent}. Research suggests that transformer-based models progressively encode increasingly complex semantic information across different layers, with hidden state evolution representing a physical correspondence to incremental reasoning \cite{peters2018dissecting,tenney2019bert,jawahar2019does,chen2020mixtext}. Thus, we hypothesize that AlphaFold3's diffusion transformer, through its Chain of Diffusion Embeddings (CODE), progressively models inter-residue relationships that directly correspond to the backbone constraints and energy biases central to topological frustration. Each transformer block refines the representation of molecular connectivity patterns, with the cumulative trajectory encoding the complexity of topological constraints present in the target structure. This connection is theoretically motivated by the physical correspondence between folding pathway complexity and representational refinement difficulty.  The rate of change in trajectory magnitude reflects energy landscape ruggedness from a topological perspective: smaller magnitude changes correspond to smoother transitions through conformational space, while larger changes indicate more constrained, difficult-to-navigate pathways. This provides a complementary assessment to energetic frustration metrics, enabling more comprehensive evaluation of structural plausibility.

Building upon the established connection between latent embedding trajectories and topological frustration, we then formalize the Chain of Diffusion Embedding (CODE) under Alphafold3's diffusion transformer $f$. Given that the diffusion transformer consists of $L$ hidden layers, we can partition $f$ into a series of ordered submodules:
\begin{equation}
    f = f_{head} \circ f_l \circ \cdots \circ f_1 \circ f_{0}
\label{f definition}
\end{equation}
In Eq.\ref{f definition}, $f_{\text{head}} : \mathbb{R}^d \rightarrow \mathbb{R}^{3}$ is the final coordinate prediction layer, $f_0 : \mathbb{R}^d \rightarrow \mathbb{R}^d$, is the diffusion conditioning module that receives the embedding from the noise coordinates and pairformer output. $f_l$ ($1 \leq l \leq L$) : $\mathbb{R}^d \rightarrow \mathbb{R}^d$ is the intermediate transformer block. Here $d$ is the embedding dimensions. Given a sequence $\mathbf{x}$ with length $N$ as input to $f$, when AlphaFold3 generates its ultimate structural prediction at the final denoising timestep, we denote the $i$-th residue's hidden representation at layer $l$ as
$\mathbf{z}^i_l$. Following the definitions in earlier research \cite{ren2022out,wang2024embedding}, we define the average embedding at layer $l$ as:
\begin{equation}
    \mathbf{h}_l^{diff} = \frac{1}{N} \sum_{i=1}^{N} \mathbf{z}^i_l
\end{equation}

\noindent which represents the $l$-th global structural representation in the diffusion process. Then, the CODE is expressed as a progressive chain $\mathcal{H}$ of all structural hidden states formalized as follows:
\begin{equation}
    \mathcal{H}^{diff} = \underbrace{\mathbf{h}_0^{diff}}_{\text{Condition}} \rightarrow \underbrace{\mathbf{h}_1^{diff} \rightarrow \cdots \rightarrow \mathbf{h}_l^{diff} \rightarrow \cdots \rightarrow \mathbf{h}_{L-1}^{diff}}_{\text{Intermediate Hidden States}} \rightarrow \underbrace{\mathbf{h}_L^{diff}}_{\text{Output State}}
\end{equation}

To quantify the trajectory of diffusion embedding changes, we need to extract basic geometric information from the chain. The magnitude of change is obviously a direct feature of the path, which is used to reflect the changes in global structural information during the wandering process \cite{helland2009trajectory,rintoul2015trajectory}. We define this basic feature as the L2-norm of magnitude change between any two adjacent embedding pair:

\begin{equation}
M^{\text{diff}}(h_l, h_{l+1}) = \|h_{l+1}^{\text{diff}} -h_l^{\text{diff}}\|_2
\end{equation}

CODE needs to consider the magnitude change characteristics of the entire change trajectory. Specifically, it sums the magnitude of each adjacent embedded change and takes the average:

\begin{equation}
    \text{CODE}(\mathcal{H}^{diff}) = \frac{1}{L} \sum_{l=0}^{L} \frac{M^{\text{diff}}(h_l, h_{l+1})}{Z^{\text{diff}}}
\end{equation}

\begin{equation}
    Z^{\text{diff}} = \|h_{L}^{\text{diff}} - h_0^{\text{diff}}\|_2
\end{equation}

To mitigate the potential sample bias, range scaling factor $Z^{diff}$ is introduced for the following reasons: If the input and output of one sample are far apart in the latent space, its trajectory naturally has a longer wandering distance. By setting scaling factors, we convert the absolute magnitude changes of each adjacent state pair into relative changes, specifically,
the changes of $(h_l, h_{l+1})$ is relative to the changes of the input-output states $(h_0, h_L)$, thereby avoiding measurement noise caused by inherent differences between samples.

\subsection{Confidence Score Computing}
The prior work \cite{roney2022state} pointed out that Alphafold has learned a certain folding energy function. They hypothesized that AlphaFold’s folding mechanism relies on an initial conformation estimate from the MSA and recycling mechanisms, enabling rapid localization of the energy landscape’s local minimum. In the second step, the structural module iteratively scores the initial conformation to yield a reliable predicted structure. Thus, we believe the AlphaFold partially captures energy frustration by predicting multiple confidence metrics learned during training, accounting for overall structural rationality.

In this paper, we mainly focus on two of them:  predicted local distance difference (pLDDT) and predicted template modeling score (pTM). The pLDDT provides a per-atom confidence estimate. For atom $a$, pLDDT is calculated as:
\begin{equation}
    pLDDT_a = \frac{1}{|R|} \sum_{m \in R} \frac{1}{4} \sum_{c \in \{0.5, 1, 2, 4\}}[|d_{am}^{\text{pred}} - d_{am}^{\text{true}}| < c \text{ Å}]
\end{equation}

\noindent where $d_{am}^{\text{pred}}$ and $d_{am}^{\text{true}}$ are the predicted and true distances between atoms $l$ and $m$, respectively. The set $R$ contains atoms satisfying some criteria which can be found in earlier research \cite{abramson2024accurate} in detail.

The pTM provides a global confidence assessment by approximating the TM-score through predicted alignment errors. The method introduces a pairwise error matrix $e_{ij}$, which captures the positional error of the C$_\alpha$ atom of residue $j$ when predicted and true structures are aligned using the backbone frame of residue $i$. The error distribution is discretized into 64 bins covering the range from 0 to 31.5 Å with 0.5 Å bin width. The pairwise error $e_{ij}$ is computed as a linear projection of the pair representation $z_{ij}$ followed by softmax normalization. Using the predicted error matrix, pTM is approximated as:
\begin{equation}
    \text{pTM} = \max_i \frac{1}{N_{\text{res}}} \sum_j \mathbb{E}[t(e_{ij})]
\end{equation}

\noindent where the expectation is taken over the probability distribution defined by $e_{ij}$, $N_{res}$ represents the residue number and $t$ represents the TM-score transformation function. Details can be found in the prior study \cite{jumper2021highly}.

Aggregating both pLDDT and pTM, Boltz1 proposed Confidence Score to rank predicted structures:

\begin{equation}
       \text{Confidence Score} = 0.8 \times \text{pLDDT} + 0.2 \times \text{pTM}
\end{equation}

The Confidence Score has a range of [0,1], where higher values indicate higher confidence.

\subsection{Reweighting CODE and Confidence Score}
Different biological systems require distinct weighting of energetic and topological frustrations due to their inherent structural and functional characteristics. The relative importance of energy frustration and topology frustration depends on the folding pathway complexity: fast-folding proteins with simple topologies may be primarily limited by energetic conflicts, while proteins with complex structures or those requiring chaperone assistance face substantial topological barriers. Therefore, we adaptively weight two components based on the specific structural class, folding mechanism, and environmental context of the target system.

To address this problem, we proposed CONFIDE (CONFIdence integrated chain of Diffusion Embedding), which dynamically integrates Confidence Score and CODE to comprehensively consider folding conformations from the perspectives of energy frustration and topological frustration. We formalize the dynamic weight optimization process for CONFIDE as follows. Given a structure prediction task, let $\mathbf{s}_{\text{code}} \in \mathbb{R}$ denote the CODE and $\mathbf{s}_{\text{conf}} \in \mathbb{R}$ represent the Confidence Scores. The CONFIDE score is computed as a weighted linear combination:
\begin{equation}
    \text{CONFIDE}(\mathbf{s}_{\text{code}}, \mathbf{s}_{\text{conf}}; w_1, w_2) = w_1 \cdot \mathbf{s}_{\text{code}} + w_2 \cdot \mathbf{s}_{\text{conf}}
\end{equation}

To establish optimal weighting that maximizes predictive power for structural accuracy, we perform a systematic grid search over the discrete weight space $\mathcal{W} = \{-10, -9, \ldots, 9, 10\}^2$. For each weight configuration $(w_1, w_2) \in \mathcal{W}$, we evaluate the Spearman rank correlation between the composite CONFIDE scores and RMSD values across the dataset $\mathcal{D}$:
\begin{equation}
   \rho_{w_1,w_2} = \text{Spearman}\left(\{\text{CONFIDE}(\mathbf{s}^{(i)}_{\text{code}}, \mathbf{s}^{(i)}_{\text{conf}}; w_1, w_2)\}_{i \in \mathcal{D}}, \{\text{RMSD}^{(i)}\}_{i \in \mathcal{D}}\right) 
\end{equation}

The optimal weight configuration is determined through:
\begin{equation}
    (w_1^*, w_2^*) = \arg\max_{(w_1,w_2) \in \mathcal{W}} \rho_{w_1,w_2}
\end{equation}

This adaptive weighting mechanism enables CONFIDE to dynamically balance the complementary information from topology frustration and energy frustration, automatically discovering the optimal linear combination that best captures the relationship.

\section{Discussion}\label{discussion}
We introduce CODE, a latent-space trajectory–based metric, and CONFIDE, an integrated score that jointly quantify topological and energetic frustrations underlying unreliable AlphaFold3 predictions. By reframing diffusion embeddings as structural “reasoning paths,” CODE captures folding pathway constraints overlooked by conventional confidence metrics such as pLDDT. Our investigations highlight CODE and CONFIDE’s ability to markedly enhance the reliability of data-driven biomolecular structure prediction.

As unsupervised, plug-and-play tools, CODE and CONFIDE demonstrate versatility across biochemical contexts. CONFIDE consistently outperforms pLDDT, showing substantial gains in hallucination detection, flexible protein modeling, and RNA aptamer screening, while CODE-guided optimization improves binder design by increasing success rates and enhancing conserved secondary-structure content without sacrificing diversity. In predictive tasks such as affinity estimation and drug-resistance profiling, CONFIDE achieves strong correlations with experimental measures (e.g., up to 0.89 in BTK inhibitor resistance prediction), underscoring its value for robust and interpretable structural modeling across applications central to drug discovery.

By revealing biophysical constraints embedded within deep structure predictors, CODE and CONFIDE provide a scalable, mechanistically interpretable toolkit with broad applications. Paired with AlphaFold3, they can generate high-quality structural datasets across biomolecular modalities, fueling advances in diffusion-based predictors, inverse-folding models, multi-conformational design, and cryptic-pocket identification. Ultimately, we envision CODE and CONFIDE as a new paradigm for unsupervised structural evaluation—accelerating progress in computational structural biology, deepening our understanding of protein folding, and driving innovations in AI-guided therapeutic discovery.

\backmatter




\section*{Declarations}

\subsection*{Author Contributions Statement}
Z.G. and C.G. conceived the idea, developed the theoretical formalism, constructed the model and conducted experiments, and wrote the manuscript; M.H., S.S. and X.W. prepared the data; H.C. helped write and revise the manuscript; J.Z. investigated the Drug resistance prediction; Z.L. verified the analytical methods; X.Y. revised manuscript and provided computing resources; C.H., C.G. and P.H. supervised the project and revised the manuscript. All authors discussed the results and contributed to the final manuscript.

\subsection*{Competing Interests Statement}
The authors declare no competing interests.

\subsection*{Data and Code Availability}
This study utilizes 13 datasets.  The topological frustration dataset is accessible from \url{https://pubs.acs.org/doi/10.1021/ja049510%2B}. The ternary complex structure prediction dataset is available from \url{https://github.com/NilsDunlop/PROTACFold}. The flexible and rigid protein datasets can be found at \url{https://www.nature.com/articles/s41592-023-01831-0#data-availability} and \url{https://www.nature.com/articles/s41467-023-36699-3#data-availability}. The PDB test and CASP 15 datasets is accessible from \url{https://drive.google.com/file/d/1JvHlYUMINOaqPTunI9wBYrfYniKgVmxf/view}. For data resources of binder design, please refer to \url{https://github.com/yehlincho/BoltzDesign1}. The enzyme site dataset is available from \url{https://www.nature.com/articles/s41467-024-51511-6#data-availability}. For the drug-resistant mutation dataset, please refer to \url{https://www.nejm.org/doi/full/10.1056/NEJMoa2114110} for detailed PDB ID. For affinity data of sEH 2009 and 2010 versions, please refer to \url{https://doi.org/10.1101/2025.04.07.647682}. The RNA aptamer dataset is accessible from \url{https://github.com/Graph-and-Geometric-Learning/Frame-Averaging-Transformer/tree/main/dataset/aptamer/raw}. All code for structure prediction and the anlysis of CODE and CONFIDE can be accessed at \url{https://github.com/zjgao02/CONFIDE.git}.



\renewcommand\thesection{S\arabic{section}}
\renewcommand{\figurename}{Supplementary Fig.}
\renewcommand{\tablename}{{Supplementary Table}}

\setcounter{section}{0}
\setcounter{figure}{0}
\setcounter{table}{0}

\section{Dataset Description}\label{secA1}

\subsection{Datasets on topological frustration}\label{topo_dataset}
We utilized a database of energetically unfrustrated single domain proteins for our analysis \cite{chavez2004quantifying}. This database represents the extrapolation of the minimal (energetic) frustration principle to the limit of completely unfrustrated protein-like chains. By studying the folding landscape in this  energetically unfrustrated protein world, we were able to concentrate on features determined solely by backbone topology (i.e., configuration entropy). The entropic roughness across different proteins in this dataset provided a clean explanation for the different folding scenarios experimentally detected: bottlenecks in configuration entropy were identified along the folding routes of slow folding proteins, whereas the smooth entropy landscapes associated with the fastest folders left more room for energetic perturbations (sequence dependence) to shape the minimal free energy pathways to the native state.

\begin{figure}[htbp]
    \centering
    \includegraphics[width=1\textwidth]{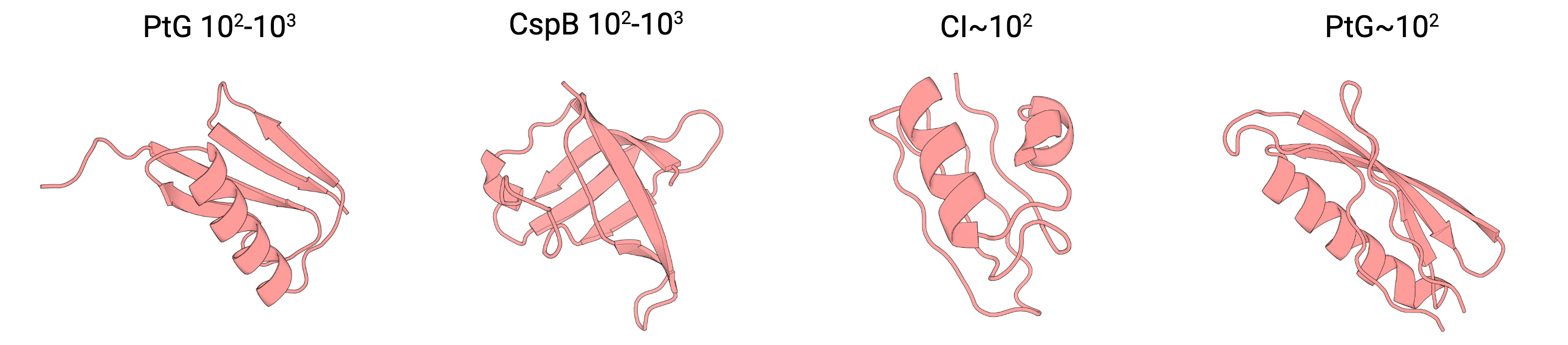}
    \caption{The predicted structures of the remaining four proteins in the main text.}
    \label{topo_a1}
\end{figure}

The proteins in the database were chosen to cover a range of experimentally determined folding rates spanning over six orders of magnitude, while exhibiting diverse overall folding topologies and secondary structure compositions. We analyzed sixteen two-state, single-domain proteins, with lengths ranging from 36 to 115 residues. Table \ref{topo data} provides a summary of the structural details and folding rates of the selected proteins. Figure \ref{topo_a1} shows the predicted structures of the remaining four proteins in the main text.

\begin{table}[h]
\centering
\caption{Summary of Most Relevant Structural and Folding Features for All Two-State Folding Proteins Used in the proof-of-concept experiments on the correlation between topological frustration and CODE.}
\begin{tabularx}{\textwidth}{llXl}
\toprule
Protein & L & Structural Information & $k_f$ (s$^{-1}$) \\
\midrule
HP36 & 36 & All helical; smallest naturally occurring, independently folding protein domain & $\sim 10^5$  \\
$\lambda_{6-85}$ & 80 & Five-helix bundle & $10^4 - 10^5$ \\
Psbd & 43 & Very small three-helix bundle & $\sim 10^4$  \\
N-L9 & 56 & Three-stranded antiparallel $\beta$-sheet sandwiched between two helices & $\sim 10^3$ \\
CspB & 67 & Small $\beta$-barrel & $10^2 - 10^3$  \\
PtG & 56 & Four-stranded $\beta$-sheet spanned by an $\alpha$-helix (similar to PtL) & $10^2 - 10^3$ \\
CI2 & 64 & Six-stranded $\beta$-sheet packed against an $\alpha$-helix & $\sim 10^2$ \\
PtL & 61 & Four-stranded $\beta$-sheet spanned by an $\alpha$-helix (similar to PtG) & $\sim 10^2$  \\
Im9 & 86 & Four-helix bundle & $\sim 10^2$  \\
SH3 & 57 & Two antiparallel $\beta$-sheets orthogonally packed & $10 - 10^2$ \\
TI-I27 & 89 & Two antiparallel $\beta$-sheets packed against each other (Greek key topology) & $1 - 10$ \\
HPr & 85 & Three $\alpha$-helices packed against a four-stranded antiparallel $\beta$-sheet & $1 - 10$  \\
MerP & 72 & Antiparallel four-stranded $\beta$-sheet, with two helices packed on one side & $\sim 1$ \\
TWIg & 93 & Two antiparallel $\beta$-sheets packed against each other (Greek key topology) & $\sim 1$  \\
AcP & 98 & Two antiparallel $\alpha$-helices packed against a five-stranded $\beta$-sheet & $\sim 10^{-2}$  \\
P13 & 115 & Filled $\beta$-barrel & $10^{-2} - 10^{-1}$  \\
\bottomrule
\label{topo data}
\end{tabularx}
\end{table}

\subsection{Datasets on ternary complex structure} \label{ternary_dataset}
We utilized the dataset curated 62 PROTAC-mediated ternary and binary complexes from the PDB \cite{dunlop2025predicting}. Each structure contains experimentally resolved coordinates for the complete PROTAC molecule--including warhead, linker, and E3 ligase binding moiety--as well as at least one of either the protein of interest (POI) or the E3 ubiquitin ligase. This collection represents all publicly available, fully crystallized PROTAC ternary and binary complexes as of May 2024. Structures with partially resolved PROTACs, such as those capturing solely the warhead component, were identified but excluded from the dataset.

The completeness of PROTAC crystallization was validated through systematic review of each structure's associated publication, enhanced by computational analysis tools. To analyze the complete ternary complex interaction, we finally retained 48 samples with complete POI, E3 and ligand. Similarly, we also cleaned out 33 molecular glue samples. These samples were finally processed into YAML files and input into Boltz1 for structure prediction without MSA. In order to better characterize ligands, we used CCD format input for ligands.

\subsection{Datasets on flexible and rigid protein structure} \label{flexible_dataset}
Flexible proteins are disordered proteins from liquid-liquid phase separation, obtained from the CD-CODE platform. This platform provides a wide range of protein sequences related to biomolecular condensates, including 223 Diver LLPS protein sequences. We randomly selected 29 of these sequences as a validation dataset. Rigid proteins were obtained from the rigid and MOAD PDB data sets of negative samples constructed using pocketminer \cite{meller2023predicting}. For details, see Supplementary Data 1 in \cite{meller2023predicting}.

\subsection{Datasets on PDB test and CASP15}\label{casp_dataset}
We evaluate the performance of the model on two benchmarks: the diverse test set of recent PDB structures  and CASP15, the last community-wide protein structure prediction competition where for the first time RNA and ligand structures were also evaluated \cite{das2023assessment,robin2023assessment}. These benchmarks feature a diverse range of structures, including protein complexes, nucleic acids, and small molecules, making them ideal for evaluating models like Boltz-1 that predict the structure of various biomolecules.

For PDB test, We followed the dataset processing strategy of the Boltz1 benchmark, and the final test set size was 593. The detailed processing process can be found in section 2.2 in \cite{wohlwend2024boltz}. 

For CASP15, the benchmark processing methodology from Boltz-1 was adopted to extract all competing targets using the following filtering criteria: (1) targets that were not withdrawn from the competition, (2) targets with associated PDB identifiers to obtain ground truth crystal structures, (3) targets where the stoichiometry information matched the number of provided chains, and (4) targets with total residue counts below 2000. This filtering process resulted in 76 structures. For the test dataset, structures containing covalently bound ligands were excluded since the current version of the Chai-1 public repository does not support configuration of these interactions. Additionally, for both datasets, structures that exceeded memory limitations or failed due to other technical issues on A100 80GB GPUs were removed from consideration. Following these preprocessing and filtering steps, the final evaluation datasets comprised 66 structures for CASP15 and 541 structures for the test set, ensuring computational feasibility and methodological consistency across all evaluated prediction methods.

\subsection{Datasets on enzyme site identification}\label{enzyme_dataset}

We used the test dataset proposed in \cite{wang2024multi} for enzyme site identification. The test dataset contains 894 data, including the enzyme catalytic reaction equation, enzyme commission number, PDB ID, amino acid sequence, site interval and label. For the label of the site interval, 0 represents the binding site, 1 represents the catalytic site, and 2 represents others.

\subsection{Datasets on BTK mutants prediction}\label{btk_dataset}
\cite{wang2022mechanisms} provided the source data, which contained experimentally measured affinity data for four BTKs with their five mutants, and the wild type. The dissociation equilibrium constants ($K_d$) of compounds to purified wild-type or mutant BTK protein are determined with the use of surface plasmon resonance technology. We treat 'no binding detected' data as 100 $\mu M$ for the correlation analysis following \cite{zheng2025alphafold3}.

\subsection{Datasets on sEH inhibitor affinity prediction}\label{seh_dataset}
Following \cite{zheng2025alphafold3}, we selected two well-characterized series of soluble epoxide hydrolase (sEH) inhibitors from the CHEMBL database, comprising compounds with a wide range of experimental IC50 and pIC50 values (pIC50 equals the negative log\textsubscript{10}IC50) but sharing common scaffolds. We utilized a dataset containing protein-ligand pairs with two versions from 2009 and 2010, comprising 25 and 50 pairs respectively. 

\subsection{Datasets on RNA aptamers screening}\label{rna_dataset}

Our analysis utilized a curated dataset of GFPapt mutants by \cite{huang2024protein}, originally derived from the work of \cite{shui2012rna}. The dataset encompasses aptamers exhibiting a broad spectrum of binding affinities, with dissociation constants (K\textsubscript{d}) ranging from 0 nM to 125 nM. For classification purposes, aptamers demonstrating K\textsubscript{d} values below 10 nM were designated as high-affinity binders and classified as positive hits for subsequent analysis.

\section{Additional Results Analysis}\label{secB1}
\subsection{Performance in PROTAC Structure Prediction}

\begin{figure}[htbp]
    \centering
    \includegraphics[width=1\textwidth]{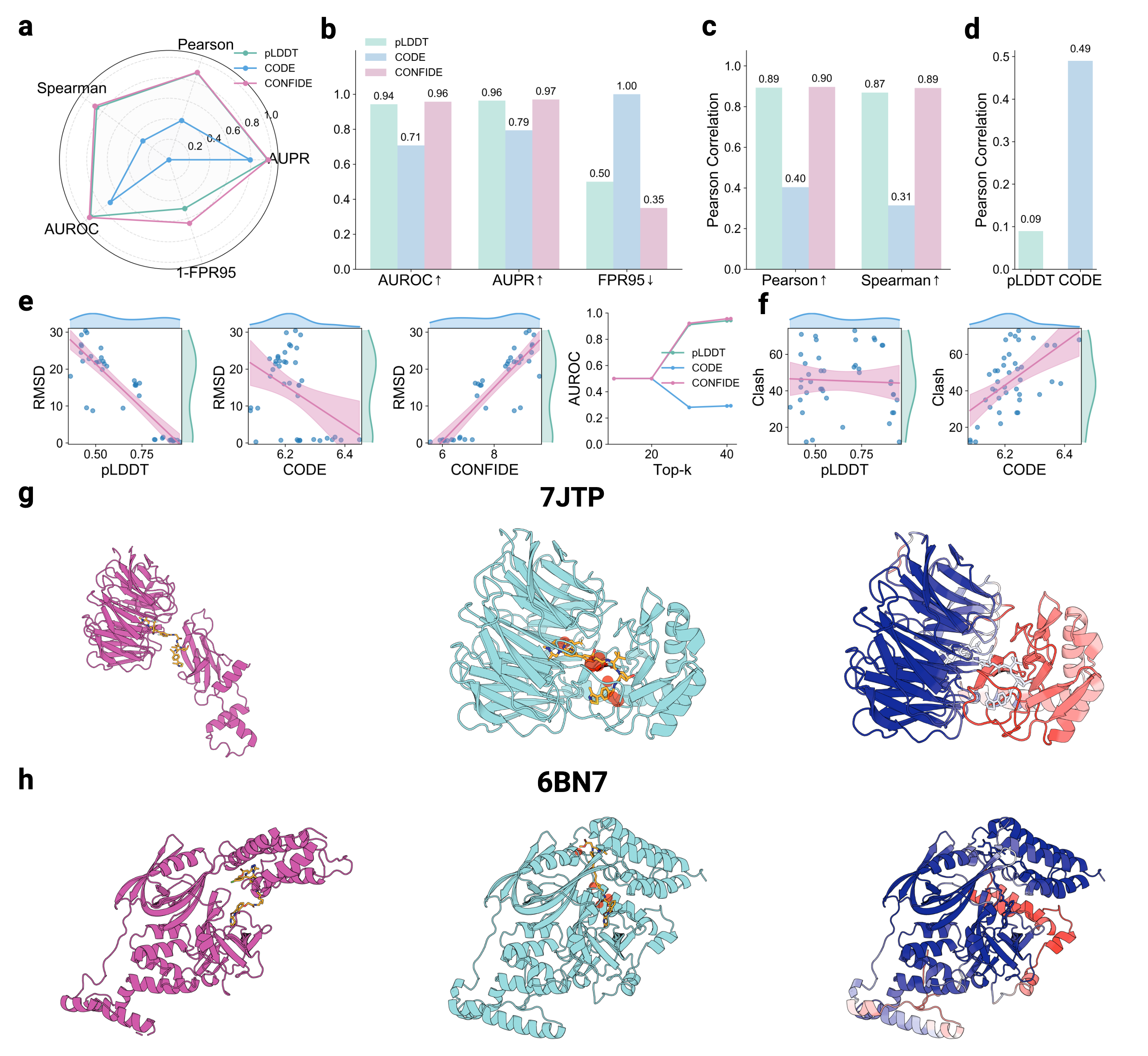}
    \caption{\textbf{Performance analysis on PROTAC structure prediction.} (a) Radar chart of the five evaluation metrics of pLDDT, CODE and CONFIDE. (b) Histogram of classification evaluation metrics of pLDDT, CODE and CONFIDE. (c) Histogram of correlation evaluation metrics of pLDDT, CODE and CONFIDE. (d) Spearman correlation between the number of atomic conflicts and pLDDT/CODE. (e) Scatter plots showing the fit and distribution of pLDDT, CODE, and CONFIDE against RMSD, including linear regression fits and 95\% confidence intervals, as well as AUROC for different classification thresholds. (f) Scatter plots of the fit and distribution of pLDDT and CODE with RMSD, incorporating linear regression fits and 95\% confidence intervals. (g-h) The true structure, predicted structure, and predicted structure colored by pLDDT of 7JTP and 6BN7 in PROTAC are shown from left to right. The red balls in the middle figure indicate atomic conflicts.}
    \label{protac_a1}
\end{figure}

In Figure \ref{protac_a1}a, we first presented an overall comparison of the performance of the three scores in classification and correlation tasks, observing that CONFIDE demonstrates an advantage over pLDDT or CODE alone. In the classification task, CONFIDE improved the AUROC by 0.02 and the FPR95 by 0.15 compared to pLDDT which is shown in Figure \ref{protac_a1}b. In the Pearson correlation analysis, CONFIDE reached 0.9, which is 0.50 higher than using CODE alone and 0.01 higher than pLDDT. In the Spearman correlation coefficient, CONFIDE reached 0.89, which is 0.58 higher than using CODE alone and 0.02 higher than pLDDT, which is shown in Figure \ref{protac_a1}c. In Figure \ref{protac_a1}e, we displayed the distribution of the three scores across all samples along with their fitted curves, as well as the trend of AUROC changes with different top K selections.

As in our in-depth analysis of molecular glue in the main text, we also demonstrated the powerful ability of CODE to recognize topological frustration caused by atomic conflicts in the PROTAC complex. Here we presented two representative PROTAC as a case study (PDB ID: 7JTP and 6BN7) in Figure \ref{protac_a1}g and Figure \ref{protac_a1}h. The leftmost image shows the true crystal structure. The middle image depicts the Boltz1-predicted structure, with atomic clash regions marked by red spheres. The rightmost image colors the predicted structure based on pLDDT scores, with blue indicating high confidence and red indicating low confidence. It was observed that pLDDT can, to some extent, reflect the quality of structural folding, particularly in assigning notably low scores to disordered or loop regions, indicating that pLDDT can capture this type of energy frustration. However, pLDDT is ineffective in detecting atomic clashes. In contrast, atomic clashes significantly increase topological frustration, enabling CODE to capture such structural prediction inaccuracies.

In addition, we statistically quantified CODE's advantages. Figure \ref{protac_a1}d shows the Pearson correlation coefficients of pLDDT and CODE with the number of atomic clashes, where CODE achieved a correlation of 0.49, a 444\% improvement over pLDDT’s 0.09. Figure \ref{protac_a1}f illustrates the specific distribution and fitted curves. These significant improvements demonstrate that CODE captures critical complementary information related to topological frustration, providing substantial benefits for evaluating the structural quality of ternary or even higher-order complexes.

\subsection{Performance in Rigid Structure Prediction}\label{srigid}

\begin{figure}[htbp]
    \centering
    \includegraphics[width=1\textwidth,height=0.3
    \textheight]{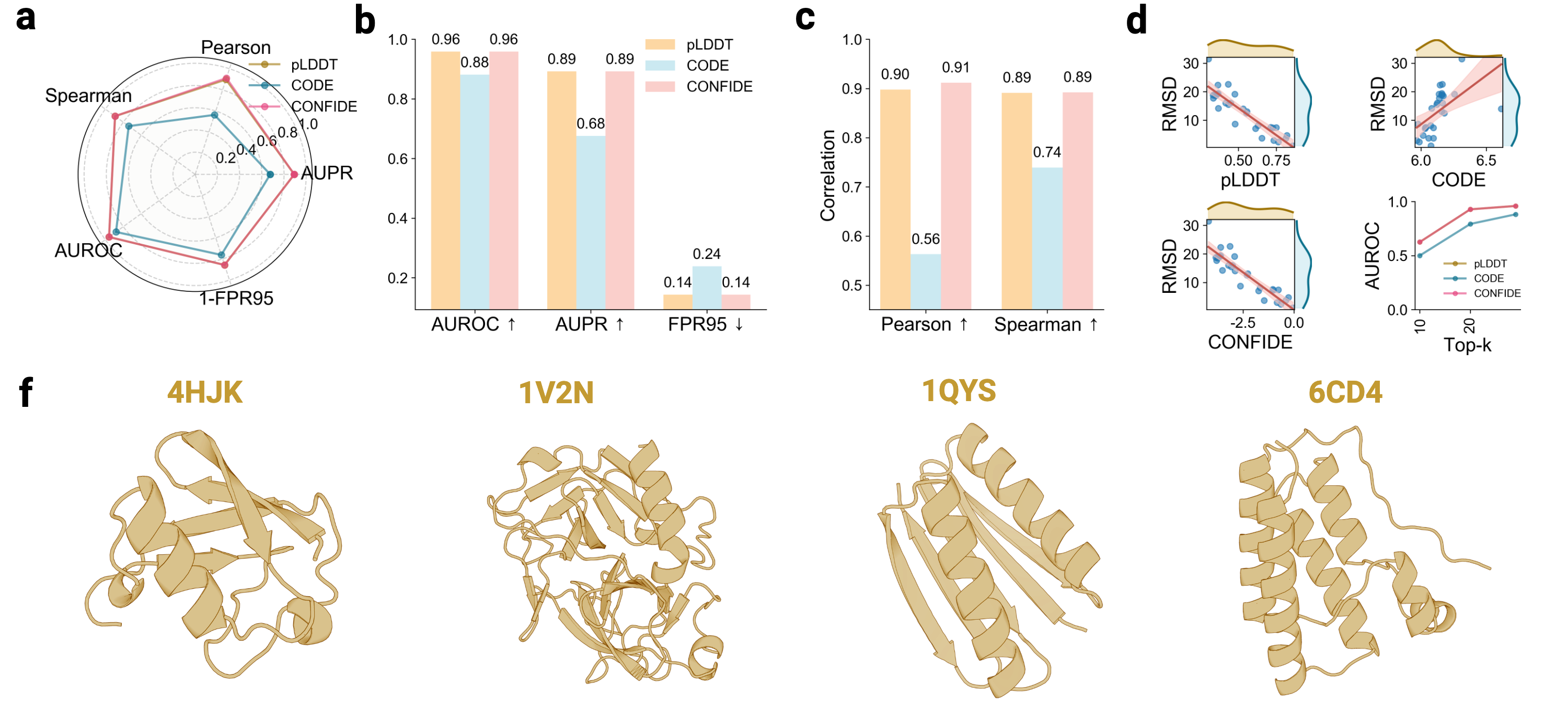}
    \caption{\textbf{Performance analysis on rigid protein structure prediction.} (a) Radar chart of the five evaluation metrics of three scores (pLDDT, CODE and CONFIDE) on rigid protein. (b) Histogram of classification evaluation metrics of three scores on rigid protein. (c) Histogram of correlation evaluation metrics of three scores on rigid protein. (d) Scatter plots of the fit and distribution of three scores on flexible proteins with RMSD, including linear regression fits and 95\% confidence intervals, as well as AUROC for different classification thresholds. (e) The structures of four rigid proteins.}
    \label{rigid}
\end{figure}

In Figure \ref{rigid}a, we first presented an overall comparison of the performance of the three scores in classification and correlation tasks. Since the structure prediction of rigid proteins is relatively simple and Boltz1 predicts well, pLDDT alone can achieve a strong correlation of 0.9. Therefore, all indicators of CONFIDE are almost on par with pLDDT and are close to saturation. Only the Pearson correlation is higher than pLDDT alone by 0.01. In Figure \ref{rigid}d, we displayed the distribution of the three scores across all samples along with their fitted curves, as well as the trend of AUROC changes with different top K selections. Figure \ref{rigid}e shows four rigid proteins.

\subsection{Comprehensive Benchmark Evaluation on CASP15 and PDBTEST Datasets}\label{scasp}

\begin{figure}[htbp]
    \centering
    \includegraphics[width=1\textwidth,height=0.55
    \textheight]{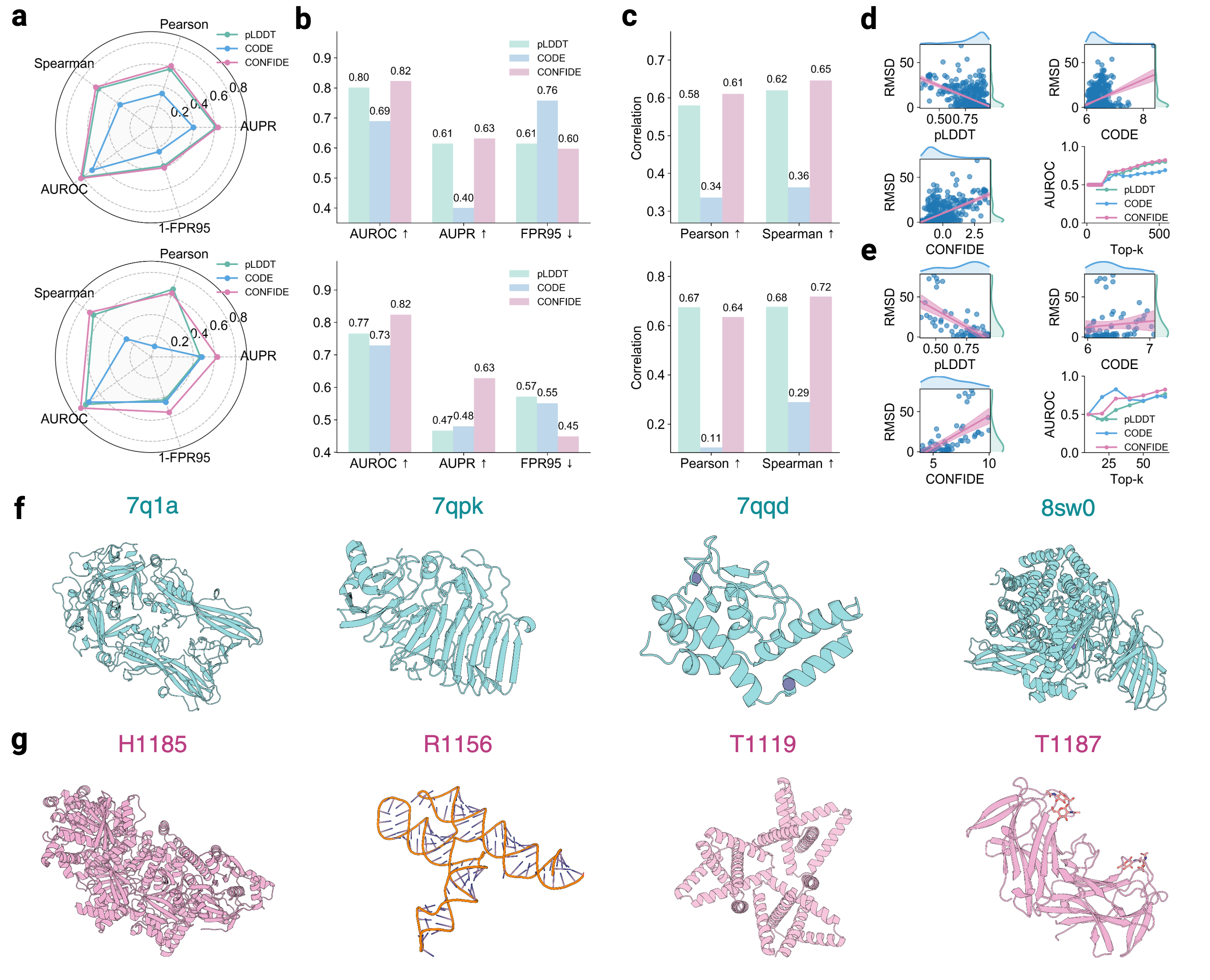}
    \caption{\textbf{Performance analysis on two recognized structure prediction benchmark PDB-test and CASP15.} (a) Radar chart of the five evaluation metrics of three scores (pLDDT, CODE and CONFIDE) on two datasets. (b) Histogram of classification evaluation metrics of three scores on two datasets. (c) Histogram of correlation evaluation metrics of three scores on two datasets. (d-e) Scatter plots of the fit and distribution of three scores on the two datasets with RMSD, including linear regression fits and 95\% confidence intervals, as well as AUROC for different classification thresholds. (f) Four representative structures predicted by Boltz1 on PDB-test. (g) Four representative structures predicted by Boltz1 on CASP15.}
    \label{benchmark}
\end{figure}

Not only do CODE and CONFIDE perform amazingly well on challenging datasets, to assess their general capabilities, we also evaluate performance on two benchmark datasets: a diverse test set containing recent PDB structures (details are provided in the supplementary material) \cite{burley2017protein}; and CASP15 (the community-wide protein structure prediction competition), which for the first time evaluated structures of RNA and ligands \cite{elofsson2023progress,das2023assessment}. Both benchmark datasets contain a very diverse set of structures, including protein complexes, nucleic acids, and small molecules, making them excellent testbeds for evaluating models that can predict the structure of arbitrary biomolecules, such as Boltz-1.

In Figure \ref{benchmark}a, we presented an overall comparison of the performance of the three scores on the PDB test and CASP 15 datasets in classification and correlation tasks. In Figure \ref{benchmark}b and Figure \ref{benchmark}c, for the PDB test dataset, in the classification task, CONFIDE improved the AUROC by 0.02 and the AUPR by 0.02 compared to pLDDT. In terms of Pearson correlation, CONFIDE achieved 0.61, surpassing pLDDT by 0.03. For Spearman correlation, CONFIDE reached 0.65, exceeding pLDDT by 0.03. For the CASP 15 dataset, in the classification task, CONFIDE achieved an AUROC of 0.82, a 0.05 improvement over pLDDT’s 0.77. In AUPR, CONFIDE reached 0.63, a 34\% improvement compared to pLDDT’s 0.47. In FPR95, CONFIDE achieved 0.45, a 21\% improvement over pLDDT’s 0.57. In Pearson correlation, CONFIDE and pLDDT were nearly equivalent, possibly due to CODE’s limited understanding of nucleic acid and small molecule folding. In Spearman correlation, CONFIDE reached 0.72, surpassing pLDDT by 0.04. In Figure \ref{benchmark}d, we displayed the distribution of the three scores across all samples along with their fitted curves, as well as the trend of AUROC changes with different top K selections. In Figures In Figure \ref{benchmark}f and Figure \ref{benchmark}g, we presented representative structures from the PDB test and CASP 15 datasets, respectively.

\section{Evaluation Metric Description}\label{metric}
To comprehensively evaluate the performance of pLDDT, CODE, and CONFIDE in different application scenarios, we use six metrics. Each metric provides a unique perspective on the classification or correlation analysis results, ensuring a comprehensive and balanced analysis of different use cases.
\begin{itemize}
\item \textbf{Area Under the Receiver Operating Characteristic Curve (AUROC)}
\[
\mathrm{AUROC} 
\:=\:
\int_0^1 \mathrm{TPR}(\mathrm{FPR}^{-1}(t)) \, dt
\:=\:
\int_0^1 \frac{TP}{TP + FN} \, d\left(\frac{FP}{FP + TN}\right),
\]
where TPR is the true positive rate and FPR is the false positive rate. AUROC measures the area under the ROC curve, providing a single scalar value representing the model's discriminative ability across all classification thresholds.

\item \textbf{Area Under the Precision-Recall Curve (AUPR)}
\[
\mathrm{AUPR} 
\:=\:
\int_0^1 \mathrm{Precision}(\mathrm{Recall}^{-1}(r)) \, dr
\:=\:
\int_0^1 \frac{TP}{TP + FP} \, d\left(\frac{TP}{TP + FN}\right).
\]
AUPR quantifies the area under the precision-recall curve, which is particularly useful for imbalanced datasets as it focuses on the positive class performance and is less influenced by the large number of true negatives.

\item \textbf{False Positive Rate at 95\% True Positive Rate (FPR95)}
\[
\mathrm{FPR95} 
\:=\:
\frac{FP}{FP + TN} \Big|_{\mathrm{TPR} = 0.95},
\]
where the threshold is set such that the true positive rate equals 95\%. FPR95 measures the false positive rate when the model achieves 95\% sensitivity, providing insight into the trade-off between sensitivity and specificity at high recall levels.

\item \textbf{Pearson Correlation Coefficient}
\[
\mathrm{Pearson} 
\:=\:
\frac{\sum_{i=1}^n (x_i - \bar{x})(y_i - \bar{y})}{\sqrt{\sum_{i=1}^n (x_i - \bar{x})^2} \sqrt{\sum_{i=1}^n (y_i - \bar{y})^2}},
\]
where $x_i$ and $y_i$ are the predicted and actual values, and $\bar{x}$ and $\bar{y}$ are their respective means. The Pearson correlation measures the linear relationship between predicted and actual values, ranging from -1 to 1.

\item \textbf{Spearman Rank Correlation Coefficient}
\[
\mathrm{Spearman} 
\:=\:
1 - \frac{6\sum_{i=1}^n d_i^2}{n(n^2 - 1)},
\]
where $d_i$ is the difference between the ranks of corresponding predicted and actual values, and $n$ is the number of observations. Spearman correlation assesses the monotonic relationship between variables, making it robust to outliers and non-linear relationships.

\item \textbf{Area Under the Precision-Recall Curve (AUPRC)}
\[
\mathrm{AUPRC} 
\:=\:
\int_0^1 \mathrm{Precision}(\mathrm{Recall}^{-1}(r)) \, dr
\:=\:
\int_0^1 \frac{TP}{TP + FP} \, d\left(\frac{TP}{TP + FN}\right).
\]
AUPRC is identical to AUPR and measures the area under the precision-recall curve, providing a comprehensive evaluation of model performance especially for imbalanced datasets where positive instances are rare.

\item \textbf{Recall}
\[
\mathrm{Recall} 
\:=\:
\frac{TP}{TP + FN}.
\]
Also known as sensitivity or true positive rate, recall measures the proportion of actual positive instances that are correctly identified by the model. High recall indicates the model's ability to detect most positive cases, which is crucial in applications where missing positive instances has serious consequences.
\end{itemize}

\section{Structural Prediction and Calculation Details}\label{compute}
Due to copyright restrictions on alphafold3, our structural inference model uses the open-source Boltz1 model. To adhere to the theoretical assumptions in the paper and to facilitate large-scale structural inference, we employed an MSA-free inference mode. After obtaining the output of Boltz1, we used the confidence score in the output JSON file to reflect energy frustration. The confidence score is calculated as $0.8 * complex_{plddt} + 0.2 * iptm$ (ptm for single chains). CODE is calculated in the hidden layers of Boltz1, and the embeddings of each layer are saved during forward inference, with a dimension of 768. After obtaining the CODE and confidence score, we selected the optimal combination coefficient based on the Spearman coefficient, or we could simply add them together. All experiments were conducted on four NVIDIA A800 80G GPUs.


\bibliography{sn-bibliography}

\end{document}